%%%%%%%%%%%%%%%%%%  tex macros for preprints, cm version %%%%%%%%%%%%%%
%         (P. Ginsparg <ginsparg@lanl.gov>, last updated 7/94)
%         hypertex extensions (still provisional), 7/26/94
%	  Some modifications by C.R.Mafra, 2012

%comment out this line to restore non-hyper functionality
%\input hyperbasics
%
\def\unredoffs{}
\tolerance=1000\hfuzz=2pt
\catcode`\@=11 % This allows us to modify PLAIN macros.
\ifx\hyperdef\UNd@FiNeD\def\hyperdef#1#2#3#4{#4}\def\hyperref#1#2#3#4{#4}\def\href#1#2{#2}\fi
\magnification=1200\unredoffs\baselineskip=16pt plus 2pt minus 1pt
\def\Date#1{\vfill\leftline{#1}\tenpoint\supereject%
\footline={\hss\tenrm\hyperdef\hypernoname{page}\folio\folio\hss}}%
% (restores pagenumbers)

%%%%%% Hour:Minute %%%%%%%%%%%%%%%%%
{\count255=\time\divide\count255 by 60 \xdef\hourmin{\number\count255}
 \multiply\count255 by-60\advance\count255 by\time
 \xdef\hourmin{\hourmin:\ifnum\count255<10 0\fi\the\count255}
}
\def\date{\number\day.\number\month.\number\year\ at \hourmin}

%%%%%%%%%%%% Draft mode %%%%%%%%%%%%%
% puts date/time on each page in big mode, writes labels in margins

% use \nolabels to get rid of eqn, ref, and fig labels in draft mode
\def\nolabels{\def\wrlabeL##1{}\def\eqlabeL##1{}\def\reflabeL##1{}}
\def\writelabels{\def\wrlabeL##1{\leavevmode\vadjust{\rlap{\smash%
{\line{{\escapechar=` \hfill\rlap{\sevenrm\hskip.03in\string##1}}}}}}}%
\def\eqlabeL##1{{\escapechar-1\rlap{\sevenrm\hskip.05in\string##1}}}%
\def\reflabeL##1{\noexpand\llap{\noexpand\sevenrm\string\string\string##1}}}
\nolabels

% tagged sec numbers
\global\newcount\secno \global\secno=0
\global\newcount\meqno \global\meqno=1
\def\s@csym{}

%%%%%%%%% Section %%%%%%%%%%%%%
\def\newsec#1\par{\global\advance\secno by1%
{\toks0{#1}\message{(\the\secno. \the\toks0)}}%
\global\subsecno=0\eqnres@t\let\s@csym\secsym\xdef\secn@m{\the\secno}\noindent
{\bf\hyperdef\hypernoname{section}{\the\secno}{\the\secno.} #1}%
\writetoca{{\string\hyperref{}{section}{\the\secno}{\bf \the\secno\quad}} {\bf #1}}\par%
\nobreak\medskip\nobreak\noindent\ignorespaces}
\def\eqnres@t{\xdef\secsym{\the\secno.}\global\meqno=1\bigbreak\bigskip}
\def\sequentialequations{\def\eqnres@t{\bigbreak}}\xdef\secsym{}

%%%%%%%% Subsection %%%%%%%%%%%
\global\newcount\subsecno \global\subsecno=0
\def\subsec#1\par{\global\advance\subsecno by1%
{\toks0{#1}\message{(\s@csym\the\subsecno. \the\toks0)}}%
\global\subsubsecno=0%
\ifnum\lastpenalty>9000\else\bigbreak\fi
\noindent{\it\hyperdef\hypernoname{subsection}{\secn@m.\the\subsecno}%
{\secn@m.\the\subsecno.} #1}\writetoca{\string\hskip1.45cm
{\string\hyperref{}{subsection}{\secn@m.\the\subsecno}{\secn@m.\the\subsecno.}}
{#1}}\par\nobreak\medskip\nobreak\noindent\ignorespaces}

%%%%%%% Appendix %%%%%%%%%%%%%%
\def\appendix#1#2{\global\meqno=1\global\subsecno=0\xdef\secsym{\hbox{#1.}}%
\bigbreak\bigskip\noindent{\bf Appendix \hyperdef\hypernoname{appendix}{#1}%
{#1.} #2}{\toks0{(#1. #2)}\message{\the\toks0}}%
\xdef\s@csym{#1.}\xdef\secn@m{#1}%
\writetoca{{\string\hyperref{}{appendix}{#1}{\bf {#1}\quad}} {\bf #2}}%
\par\nobreak\medskip\nobreak}

% \eqn\label{a+b=c}   gives displayed equation, numbered consecutively within sections.
% \eqnn, \eqna        define labels in advance, use \eqna\label before an eqalign and
%                     later \label a, \label b etc inside eqalign to get (2.3a), (2.3b) etc
%
\def\checkm@de#1#2{\ifmmode{\def\f@rst##1{##1}\hyperdef\hypernoname{equation}%
{#1}{#2}}\else\hyperref{}{equation}{#1}{#2}\fi}
\def\eqnn#1{\DefWarn#1\xdef #1{(\noexpand\relax\noexpand\checkm@de%
{\s@csym\the\meqno}{\secsym\the\meqno})}%
\wrlabeL#1\writedef{#1\leftbracket#1}\global\advance\meqno by1}
\def\f@rst#1{\c@t#1a\em@ark}\def\c@t#1#2\em@ark{#1}
\def\eqna#1{\DefWarn#1\wrlabeL{#1$\{\}$}%
\xdef #1##1{(\noexpand\relax\noexpand\checkm@de%
{\s@csym\the\meqno\noexpand\f@rst{##1}1}{\hbox{$\secsym\the\meqno##1$}})}
\writedef{#1\numbersign1\leftbracket#1{\numbersign1}}\global\advance\meqno by1}
\def\eqn#1#2{\DefWarn#1%
\xdef #1{(\noexpand\hyperref{}{equation}{\s@csym\the\meqno}%
{\secsym\the\meqno})}$$#2\eqno(\hyperdef\hypernoname{equation}%
{\s@csym\the\meqno}{\secsym\the\meqno})\eqlabeL#1$$%
\writedef{#1\leftbracket#1}\global\advance\meqno by1}
\def\xeqn{\expandafter\xe@n}\def\xe@n(#1){#1}
\def\xeqna#1{\expandafter\xe@n#1}
\def\eqns#1{(\e@ns #1{\hbox{}})}
\def\e@ns#1{\ifx\UNd@FiNeD#1\message{eqnlabel \string#1 is undefined.}%
\xdef#1{(?.?)}\fi{\let\hyperref=\relax\xdef\next{#1}}%
\ifx\next\em@rk\def\next{}\else%
\ifx\next#1\xeqn#1\else\def\n@xt{#1}\ifx\n@xt\next#1\else\xeqna#1\fi
\fi\let\next=\e@ns\fi\next}

\def\DefWarn#1{\ifx\UNd@FiNeD#1\else
\immediate\write16{*** WARNING: the label \string#1 is already defined ***}\fi}
%
% footnotes
\newskip\footskip\footskip14pt plus 1pt minus 1pt %sets footnote baselineskip
\def\footnotefont{\ninepoint}\def\f@t#1{\footnotefont #1\@foot}
\def\f@@t{\baselineskip\footskip\bgroup\footnotefont\aftergroup\@foot\let\next}
\setbox\strutbox=\hbox{\vrule height9.5pt depth4.5pt width0pt}
\global\newcount\ftno \global\ftno=0
\def\foot{\global\advance\ftno by1\def\foot@rg{\hyperref{}{footnote}%
{\the\ftno}{\the\ftno}\xdef\foot@rg{\noexpand\hyperdef\noexpand\hypernoname%
{footnote}{\the\ftno}{\the\ftno}}}\footnote{$^{\foot@rg}$}}
%
%
%     \ref\label{text}
% generates a number, assigns it to \label, generates an entry.
% To list the refs on a separate page,  \listrefs
%
\global\newcount\refno \global\refno=1
\newwrite\rfile
\def\ref{[\hyperref{}{reference}{\the\refno}{\the\refno}]\nref}
\def\nref#1{\DefWarn#1%
\xdef#1{[\noexpand\hyperref{}{reference}{\the\refno}{\the\refno}]}%
\writedef{#1\leftbracket#1}%
\ifnum\refno=1\immediate\openout\rfile=\jobname.refs\fi
\chardef\wfile=\rfile\immediate\write\rfile{\noexpand\item{[\noexpand\hyperdef%
\noexpand\hypernoname{reference}{\the\refno}{\the\refno}]\ }%
\reflabeL{#1\hskip.31in}\pctsign}\global\advance\refno by1\findarg}
%	horrible hack to sidestep tex \write limitation
\def\findarg#1#{\begingroup\obeylines\newlinechar=`\^^M\pass@rg}
{\obeylines\gdef\pass@rg#1{\writ@line\relax #1^^M\hbox{}^^M}%
\gdef\writ@line#1^^M{\expandafter\toks0\expandafter{\striprel@x #1}%
\edef\next{\the\toks0}\ifx\next\em@rk\let\next=\endgroup\else\ifx\next\empty%
\else\immediate\write\wfile{\the\toks0}\fi\let\next=\writ@line\fi\next\relax}}
\def\striprel@x#1{} \def\em@rk{\hbox{}}
\def\lref{\begingroup\obeylines\lr@f}
\def\lr@f#1#2{\DefWarn#1\gdef#1{\let#1=\UNd@FiNeD\ref#1{#2}}\endgroup\unskip}

\def\addref#1{\immediate\write\rfile{\noexpand\item{}#1}} %now unnecessary
\def\listrefs{\vfill\supereject\immediate\closeout\rfile\writestoppt
\baselineskip=\footskip\centerline{{\bf References}}\bigskip{\parindent=20pt%
\frenchspacing\escapechar=` \input \jobname.refs\vfill\eject}\nonfrenchspacing}
\def\startrefs#1{\immediate\openout\rfile=\jobname.refs\refno=#1}
\def\xref{\expandafter\xr@f}\def\xr@f[#1]{#1}
\def\refs#1{\count255=1[\r@fs #1{\hbox{}}]}
\def\r@fs#1{\ifx\UNd@FiNeD#1\message{reflabel \string#1 is undefined.}%
\nref#1{need to supply reference \string#1.}\fi%
\vphantom{\hphantom{#1}}{\let\hyperref=\relax\xdef\next{#1}}%
\ifx\next\em@rk\def\next{}%
\else\ifx\next#1\ifodd\count255\relax\xref#1\count255=0\fi%
\else#1\count255=1\fi\let\next=\r@fs\fi\next}
%

%
% this is ugly, but moore insists
\newwrite\ffile\global\newcount\figno \global\figno=1
\def\fig{fig.~\hyperref{}{figure}{\the\figno}{\the\figno}\nfig}
\def\nfig#1{\DefWarn#1%
\xdef#1{fig.~\noexpand\hyperref{}{figure}{\the\figno}{\the\figno}}%
\writedef{#1\leftbracket fig.\noexpand~\xfig#1}%
\ifnum\figno=1\immediate\openout\ffile=\jobname.figs\fi\chardef\wfile=\ffile%
{\let\hyperref=\relax
\immediate\write\ffile{\noexpand\medskip\noexpand\item{Fig.\ %
\noexpand\hyperdef\noexpand\hypernoname{figure}{\the\figno}{\the\figno}. }
\reflabeL{#1\hskip.55in}\pctsign}}\global\advance\figno by1\findarg}
\def\xfig{\expandafter\xf@g}\def\xf@g fig.\penalty\@M\ {}
\def\figs#1{figs.~\f@gs #1{\hbox{}}}
\def\f@gs#1{{\let\hyperref=\relax\xdef\next{#1}}\ifx\next\em@rk\def\next{}\else
\ifx\next#1\xfig #1\else#1\fi\let\next=\f@gs\fi\next}
%
%% because TeXlive 2011 is buggy wrt to tikz pictures with plain TeX..
\def\figin{\epsfcheck\figin}\def\figins{\epsfcheck\figins}
\def\epsfcheck{\ifx\epsfbox\UnDeFiNeD
\message{(NO epsf.tex, FIGURES WILL BE IGNORED)}
\gdef\figin##1{\vskip2in}\gdef\figins##1{\hskip.5in}% blank space instead
\else\message{(FIGURES WILL BE INCLUDED)}%
\gdef\figin##1{##1}\gdef\figins##1{##1}\fi}
\def\DefWarn#1{}
\def\figinsert{\goodbreak\topinsert}
\def\ifig#1#2#3{\DefWarn#1\xdef#1{fig.~\the\figno}
\writedef{#1\leftbracket fig.\noexpand~\the\figno}%
\figinsert\figin{\centerline{#3}}
\smallskip
\leftskip=20pt \rightskip=20pt
\baselineskip12pt\noindent
{{\bf Fig.~\the\figno}\ \ninepoint #2}
\medskip
\global\advance\figno by1\par\endinsert}
%%%%%%%%%%%%%%%%%%%%%%%%%%%%%%%%%%%%%%%%%%%%%%%%%%%%%%%%%
\newwrite\lfile
{\escapechar-1\xdef\pctsign{\string\%}\xdef\leftbracket{\string\{}
\xdef\rightbracket{\string\}}\xdef\numbersign{\string\#}}
\def\writedefs{\immediate\openout\lfile=label.defs \def\writedef##1{%
{\let\hyperref=\relax\let\hyperdef=\relax\let\hypernoname=\relax
 \immediate\write\lfile{\string\def\string##1\rightbracket}}}}%
\def\writestop{\def\writestoppt{\immediate\write\lfile{\string\pageno
 \the\pageno\string\startrefs\leftbracket\the\refno\rightbracket
 \string\def\string\secsym\leftbracket\secsym\rightbracket
 \string\secno\the\secno\string\meqno\the\meqno}\immediate\closeout\lfile}}
\def\writestoppt{}\def\writedef#1{}

% Section, subsection and appendix labels %
% Note that there must be a blanck line after \newsec,\subsec and before \seclab,\subseclab!
\def\seclab#1{\DefWarn#1%
\xdef #1{\noexpand\hyperref{}{section}{\the\secno}{\the\secno}}%
\writedef{#1\leftbracket#1}\wrlabeL{#1=#1}}
\def\subseclab#1{\DefWarn#1%
\xdef #1{\noexpand\hyperref{}{subsection}{\the\secno.\the\subsecno}%
{\the\secno.\the\subsecno}}\writedef{#1\leftbracket#1}\wrlabeL{#1=#1}}
\def\applab#1{\DefWarn#1%
\xdef #1{\noexpand\hyperref{}{appendix}{\secn@m}{\secn@m}}%
\writedef{#1\leftbracket#1}\wrlabeL{#1=#1}}
\newwrite\tfile \def\writetoca#1{}
\def\leaderfill{\leaders\hbox to 1em{\hss.\hss}\hfill}
% use this to write file with table of contents
\def\writetoc{\immediate\openout\tfile=\jobname.toc
   \def\writetoca##1{{\edef\next{\write\tfile{\noindent ##1
   \string\leaderfill{
% comment this line if you don't want hyperlinked page numbers on TOC
   \string\hyperref{}{page}{\noexpand\number\pageno}%
   {\noexpand\number\pageno}} \par}}\next}}
}
% and this lists table of contents on second pass
\newread\ch@ckfile
\def\listtoc{\immediate\closeout\tfile\immediate\openin\ch@ckfile=\jobname.toc
\ifeof\ch@ckfile\message{no file \jobname.toc, no table of contents this pass}%
\else\closein\ch@ckfile\centerline{\bf Contents}\nobreak\medskip%
{\baselineskip=16pt\footnotefont\parskip=0pt\catcode`\@=11\input\jobname.toc
\catcode`\@=12\bigbreak\bigskip}\fi}
\catcode`\@=12 % at signs are no longer letters
\def\tenpoint{\def\rm{\fam0\tenrm}% switch back to 10-point type
\textfont0=\tenrm \scriptfont0=\sevenrm \scriptscriptfont0=\fiverm
\textfont1=\teni  \scriptfont1=\seveni  \scriptscriptfont1=\fivei
\textfont2=\tensy \scriptfont2=\sevensy \scriptscriptfont2=\fivesy
\textfont\itfam=\tenit \def\it{\fam\itfam\tenit}\def\footnotefont{\ninepoint}%
\textfont\bffam=\tenbf \def\bf{\fam\bffam\tenbf}\def\sl{\fam\slfam\tensl}\rm}
\font\ninerm=cmr9 \font\sixrm=cmr6 \font\ninei=cmmi9 \font\sixi=cmmi6
\font\ninesy=cmsy9 \font\sixsy=cmsy6 \font\ninebf=cmbx9
\font\nineit=cmti9 \font\ninesl=cmsl9 \skewchar\ninei='177
\skewchar\sixi='177 \skewchar\ninesy='60 \skewchar\sixsy='60
\def\ninepoint{\def\rm{\fam0\ninerm}% switch to footnote font
\textfont0=\ninerm \scriptfont0=\sixrm \scriptscriptfont0=\fiverm
\textfont1=\ninei \scriptfont1=\sixi \scriptscriptfont1=\fivei
\textfont2=\ninesy \scriptfont2=\sixsy \scriptscriptfont2=\fivesy
\textfont\itfam=\ninei \def\it{\fam\itfam\nineit}\def\sl{\fam\slfam\ninesl}%
\textfont\bffam=\ninebf \def\bf{\fam\bffam\ninebf}\rm}
%
%---------------------------------------------------------------------
\hyphenation{anom-aly anom-alies coun-ter-term coun-ter-terms}

%%%%%%%%%%%%%%% Subsubsection %%%%%%%%%%%%%%%%%%%%%%%%%%%%%%%%%%%%
\global\newcount\subsubsecno \global\subsubsecno=0
\def\subsubsec#1\par{\global\advance\subsubsecno by1%
{\toks0{#1}\message{(\the\secno\the\subsecno\the\subsubsecno. \the\toks0)}}%
\ifnum\lastpenalty>9000\else\bigbreak\fi
\noindent{\it\hyperdef\hypernoname{subsubsection}{\the\secno.\the\subsecno\the\subsubsecno}%
{\the\secno.\the\subsecno.\the\subsubsecno.} #1}
%%% Add Subsubsections to Index
%\writetoca{\string\quad{\string\hyperref{}{subsubsection}{\the\secno\the\subsecno\the
%\subsubsecno}{\baselineskip=9pt\it\the\secno.\the\subsecno.\the\subsubsecno.}}
% {\baselineskip=9pt\it\ #1}}
\par\nobreak\medskip\nobreak\noindent\ignorespaces}

% Caption for inline tikzpictures
\def\DefWarn#1{}
\def\tikzcaption#1#2{\DefWarn#1\xdef#1{Fig.~\the\figno}
\writedef{#1\leftbracket Fig.\noexpand~\the\figno}%
{
\smallskip
\leftskip=20pt \rightskip=20pt \baselineskip12pt\noindent
{{\bf Fig.~\the\figno}\ \ninepoint #2}
\bigskip
\global\advance\figno by1 \par}}

% convert numbers [1-9] to upper case letters [A-I]
\def\ntoalpha#1{%
\ifcase#1%
@%
\or A\or B\or C\or D\or E\or F\or G\or H\or I
\fi
}

% Appendix label
\global\newcount\appno \global\appno=1
\def\applab#1{\xdef #1{\ntoalpha\appno}\writedef{#1\leftbracket#1}\wrlabeL{#1=#1}
\global\advance\appno by1}

% Clean up the title page definitions
\def\preprint#1 #2\par{\rightline{\vbox{\baselineskip12pt\hbox{#1}\hbox{#2}}}\vskip2cm}
% title with more than one line (note the blanck line in between)
%\title some line
%
%\tile another line
\def\title#1\par{\centerline{\bf #1}\nopagenumbers\pageno=0}
\def\author#1\par{\bigskip\bigskip\centerline{#1}}

\newcount\addressno

\def\email#1#2{\unskip$^#1$\footnote{\null}{\kern-\parindent \llap{$^#1$\hskip1pt}email: #2}}

% centermode for address lines
\def\startcenter{%
  \par
  \begingroup
  \leftskip=0pt plus 1fil
  \rightskip=\leftskip
  \parindent=0pt
  \parfillskip=0pt
}
\def\stopcenter{\endgroup}

\def\address{\bigskip%
  \ifnum\the\addressno=0\else\stopcenter\endgroup\fi
  \advance\addressno by 1%
  \begingroup
  \startcenter
  \it
  \obeylines
  \addressAux
}
\def\addressAux#1{#1}

% need to stop center mode and obeylines from address
\def\abstract{\stopcenter\endgroup\bigskip\bigskip\noindent}

% some sample definitions
\def\Dsl{\,\raise.15ex\hbox{/}\mkern-13.5mu D} %this one can be subscripted
\def\dsl{\raise.15ex\hbox{/}\kern-.57em\partial}
 
\def\boxeqn#1{\vcenter{\vbox{\hrule\hbox{\vrule\kern3pt\vbox{\kern3pt
	\hbox{${\displaystyle #1}$}\kern3pt}\kern3pt\vrule}\hrule}}}

\def\lform{\hbox{$\sqcup$}\llap{\hbox{$\sqcap$}}}
 %pound sterling

\def\half{{1\over 2}}

\def\bar{\overline}
\def\({\left(}
\def\){\right)}

 %redefine plain TeX \Im..
%shuffle product

%\owedge

% From Knuth's \pfbox macro
\def\qed{\hbox{\hskip 3pt
%\lower2pt
\vbox{\hrule\hbox to 7pt{\vrule height 7pt\hfill\vrule}
\hrule}}\hskip3pt}

% do not display overfull marks
\overfullrule=0pt\relax

\frenchspacing

% define labels in advance
\newread\instream \openin\instream= label.defs
\ifeof\instream \message{No labels in advance yet. Wait till next pass.}
\else \closein\instream \input label.defs
\fi
\writedefs

%%% References with hyperlinks to arxiv.org; both styles accepted
% Change arXiv to \arXiv ie
% [arXiv:hep-th/1234567].     --> [\arXiv:hep-th/1234567].
% [arXiv:1234.5678 [hep-th]]. --> [\arXiv:1234.5678 [hep-th]].
% Need to strip trailing [hep-th] (if present) to define valid URL
\def\arXiv:#1].{\hepthStrip#1 \nil}
\def\hepthStrip#1 #2\nil{\href{http://arxiv.org/abs/#1}{arXiv:#1 #2\unskip}].}

\input epsf
\input figflow

\def\centretable#1{ \hbox to \hsize {\hfill\vbox{
                    \offinterlineskip \tabskip=0pt \halign{#1} }\hfill} }

\preprint UU--2021--053

\title Taming the 11D pure spinor b-ghost

\author Max Guillen\email{\dagger}{max.guillen@physics.uu.se}

\address
%$^\dagger$ 
Department of Physics and Astronomy, 75108 Uppsala, Sweden

\abstract
We provide an alternative compact expression for the 11D pure spinor b-ghost by introducing a new set of negative ghost number operators made out of non-minimal pure spinor variables. Using the algebraic properties satisfied by these operators, it will be straightforwardly shown that $\{Q, b\}={P^2\over 2}$, as well as $\{b,b\} = Q\Omega$. As an application of this novel formulation, the ghost number two vertex operator will easily be obtained in a completely covariant manner from a standard descent relation, the ghost number three vertex operator will be shown to satisfy the generalized Siegel gauge condition, and the 11D supergravity two-particle superfield will be constructed in a quite simple way.

\bigskip
\bigskip
\bigskip
\bigskip
\centerline{\it Dedicated to the memory of Pedro Quiroz Santillan}
\vskip .2in
\Date {December 2022}

\newif\iffig
\figfalse
% to speed up compilation comment the following line
%\input tikz \figtrue

%\def\Box#1,#2,#3,#4,{{\cal N}^{(4)}_{#1|#2,#3,#4}\, I^{(4)}_{#1,#2,#3,#4}}
%\def\Pentagon#1,#2,#3,#4,#5,{{\cal N}^{(5)}_{#1|#2,#3,#4,#5}(\ell) I^{(5)}_{#1,#2,#3,#4,#5}}

%**************************************

\lref\psparticle{
N.~Berkovits,
``Covariant quantization of the superparticle using pure spinors,''
JHEP {\bf 09}, 016 (2001).
[arXiv:hep-th/0105050 [hep-th]].
%146 citations counted in INSPIRE as of 13 Dec 2022
}

\lref\ICTP{
	N.~Berkovits,
  	``ICTP lectures on covariant quantization of the superstring,''
	[hep-th/0209059].
	%%CITATION = hep-th/0209059%%
}

\lref\pselevenparticle{
%\cite{Guillen:2017mte}
M.~Guillen,
``Equivalence of the 11D pure spinor and Brink-Schwarz-like superparticle cohomologies,''
Phys. Rev. D {\bf 97}, no.6, 066002 (2018).
[arXiv:1705.06316 [hep-th]].
%5 citations counted in INSPIRE as of 13 Dec 2022
}

\lref\superpoincarequantization{
N.~Berkovits,
``Super Poincare covariant quantization of the superstring,''
JHEP {\bf 04}, 018 (2000).
[arXiv:hep-th/0001035 [hep-th]].
%576 citations counted in INSPIRE as of 20 Jul 2021
}

\lref\pssupermembrane{
%\cite{Berkovits:2002uc}
N.~Berkovits,
``Towards covariant quantization of the supermembrane,''
JHEP {\bf 09}, 051 (2002).
[arXiv:hep-th/0201151 [hep-th]].
%99 citations counted in INSPIRE as of 13 Dec 2022
}

\lref\neight{
M.~Cederwall,
``N=8 superfield formulation of the Bagger-Lambert-Gustavsson model,''
JHEP {\bf 09}, 116 (2008).
[arXiv:0808.3242 [hep-th]].
%39 citations counted in INSPIRE as of 13 Dec 2022
}

\lref\nsix{
M.~Cederwall,
``Superfield actions for N=8 and N=6 conformal theories in three dimensions,''
JHEP {\bf 10}, 070 (2008).
[arXiv:0809.0318 [hep-th]].
%37 citations counted in INSPIRE as of 13 Dec 2022
}

\lref\nfour{
M.~Cederwall,
``An off-shell superspace reformulation of D=4, N=4 super-Yang-Mills theory,''
Fortsch. Phys. {\bf 66}, no.1, 1700082 (2018).
[arXiv:1707.00554 [hep-th]].
%6 citations counted in INSPIRE as of 13 Dec 2022
}

\lref\pssugra{
M.~Cederwall,
``D=11 supergravity with manifest supersymmetry,''
Mod. Phys. Lett. A {\bf 25}, 3201-3212 (2010).
[arXiv:1001.0112 [hep-th]].
%50 citations counted in INSPIRE as of 13 Dec 2022
}

\lref\psborninfeld{
M.~Cederwall and A.~Karlsson,
``Pure spinor superfields and Born-Infeld theory,''
JHEP {\bf 11}, 134 (2011).
[arXiv:1109.0809 [hep-th]].
%26 citations counted in INSPIRE as of 13 Dec 2022
}

\lref\mafraone{
N.~Berkovits and C.~R.~Mafra,
``Some Superstring Amplitude Computations with the Non-Minimal Pure Spinor Formalism,''
JHEP {\bf 11}, 079 (2006).
[arXiv:hep-th/0607187 [hep-th]].
%77 citations counted in INSPIRE as of 13 Dec 2022
}

\lref\mafratwo{
H.~Gomez and C.~R.~Mafra,
``The Overall Coefficient of the Two-loop Superstring Amplitude Using Pure Spinors,''
JHEP {\bf 05}, 017 (2010).
[arXiv:1003.0678 [hep-th]].
%44 citations counted in INSPIRE as of 13 Dec 2022
}

\lref\mafrathree{
H.~Gomez and C.~R.~Mafra,
``The closed-string 3-loop amplitude and S-duality,''
JHEP {\bf 10}, 217 (2013).
[arXiv:1308.6567 [hep-th]].
%95 citations counted in INSPIRE as of 13 Dec 2022
}

\lref\rnspsone{
N.~Berkovits,
``Covariant Map Between Ramond-Neveu-Schwarz and Pure Spinor Formalisms for the Superstring,''
JHEP {\bf 04}, 024 (2014).
[arXiv:1312.0845 [hep-th]].
%6 citations counted in INSPIRE as of 13 Dec 2022
}

\lref\rnspstwo{
N.~Berkovits,
``Manifest spacetime supersymmetry and the superstring,''
JHEP {\bf 10}, 162 (2021).
[arXiv:2106.04448 [hep-th]].
%6 citations counted in INSPIRE as of 13 Dec 2022
}

\lref\maxmaor{
M.~Ben-Shahar and M.~Guillen,
``10D super-Yang-Mills scattering amplitudes from its pure spinor action,''
JHEP {\bf 12}, 014 (2021).
[arXiv:2108.11708 [hep-th]].
%9 citations counted in INSPIRE as of 13 Dec 2022
}

\lref\stieberger{
C.~R.~Mafra, O.~Schlotterer and S.~Stieberger,
``Complete N-Point Superstring Disk Amplitude I. Pure Spinor Computation,''
Nucl. Phys. B {\bf 873}, 419-460 (2013).
[arXiv:1106.2645 [hep-th]].
%194 citations counted in INSPIRE as of 13 Dec 2022
}

\lref\mafraoli{
C.~R.~Mafra and O.~Schlotterer,
``Multiparticle SYM equations of motion and pure spinor BRST blocks,''
JHEP {\bf 07}, 153 (2014).
[arXiv:1404.4986 [hep-th]].
%70 citations counted in INSPIRE as of 13 Dec 2022
}

\lref\dynamical{
N.~Berkovits,
``Dynamical twisting and the b ghost in the pure spinor formalism,''
JHEP {\bf 06}, 091 (2013).
[arXiv:1305.0693 [hep-th]].
}

\lref\xiyin{
C.~M.~Chang, Y.~H.~Lin, Y.~Wang and X.~Yin,
``Deformations with Maximal Supersymmetries Part 2: Off-shell Formulation,''
JHEP {\bf 04}, 171 (2016).
[arXiv:1403.0709 [hep-th]].
%15 citations counted in INSPIRE as of 13 Dec 2022
}

\lref\chiralmax{
M.~Guillen,
``Green-Schwarz and pure spinor formulations of chiral strings,''
JHEP {\bf 12}, 029 (2021).
[arXiv:2108.11724 [hep-th]].
%1 citations counted in INSPIRE as of 13 Dec 2022
}

\lref\bcjone{
Z.~Bern, J.~J.~M.~Carrasco and H.~Johansson,
``New Relations for Gauge-Theory Amplitudes,''
Phys. Rev. D {\bf 78}, 085011 (2008).
[arXiv:0805.3993 [hep-ph]].
%1030 citations counted in INSPIRE as of 13 Dec 2022
}

\lref\bcjtwo{
Z.~Bern, J.~J.~M.~Carrasco and H.~Johansson,
``Perturbative Quantum Gravity as a Double Copy of Gauge Theory,''
Phys. Rev. Lett. {\bf 105}, 061602 (2010).
[arXiv:1004.0476 [hep-th]].
%743 citations counted in INSPIRE as of 13 Dec 2022
}

\lref\bcjthree{
Z.~Bern, J.~J.~Carrasco, M.~Chiodaroli, H.~Johansson and R.~Roiban,
``The Duality Between Color and Kinematics and its Applications,''
[arXiv:1909.01358 [hep-th]].
}

\lref\maximalloopcederwall{
M.~Cederwall and A.~Karlsson,
``Loop amplitudes in maximal supergravity with manifest supersymmetry,''
JHEP {\bf 03}, 114 (2013).
[arXiv:1212.5175 [hep-th]].
%20 citations counted in INSPIRE as of 13 Dec 2022
}

\lref\maxnotesworldline{
M.~Guillen,
``Notes on the 11D pure spinor wordline vertex operators,''
JHEP {\bf 08}, 122 (2020).
[arXiv:2006.06022 [hep-th]].
%3 citations counted in INSPIRE as of 13 Dec 2022
}

\lref\OdaTonin{
I.~Oda and M.~Tonin,
``On the Berkovits covariant quantization of GS superstring,''
Phys. Lett. B {\bf 520}, 398-404 (2001).
[arXiv:hep-th/0109051 [hep-th]].
%65 citations counted in INSPIRE as of 20 Jul 2021
}

\lref\perturbiner{
A.~A.~Rosly and K.~G.~Selivanov,
``On amplitudes in selfdual sector of Yang-Mills theory,''
Phys. Lett. B {\bf 399}, 135-140 (1997).
[arXiv:hep-th/9611101 [hep-th]].
%68 citations counted in INSPIRE as of 13 Dec 2022
}
\lref\NMPS{
	N.~Berkovits,
	``Pure spinor formalism as an N=2 topological string,''
	JHEP {\bf 0510}, 089 (2005).
	[hep-th/0509120].
	%%CITATION = hep-th/0509120%%
}

\lref\elevendsimplifiedb{
N.~Berkovits and M.~Guillen,
``Simplified $D = 11$ pure spinor $b$ ghost,''
JHEP {\bf 07}, 115 (2017).
[arXiv:1703.05116 [hep-th]].
%3 citations counted in INSPIRE as of 13 Dec 2022
}
\lref\brinkschwarz{
L.~Brink and J.~H.~Schwarz,
``Quantum Superspace,''
Phys. Lett. B {\bf 100}, 310-312 (1981).
%435 citations counted in INSPIRE as of 13 Dec 2022
}

\lref\brinkhowe{
L.~Brink and P.~S.~Howe,
``Eleven-Dimensional Supergravity on the Mass-Shell in Superspace,''
Phys. Lett. B {\bf 91}, 384-386 (1980).
%170 citations counted in INSPIRE as of 13 Dec 2022
}

\lref\quartet{
T.~Kugo and I.~Ojima,
``Local Covariant Operator Formalism of Nonabelian Gauge Theories and Quark Confinement Problem,''
Prog. Theor. Phys. Suppl. {\bf 66}, 1-130 (1979).
}

\lref\cederwallequations{
N.~Berkovits and M.~Guillen,
``Equations of motion from Cederwall's pure spinor superspace actions,''
JHEP {\bf 08}, 033 (2018).
[arXiv:1804.06979 [hep-th]].
}

\lref\pssreview{
M.~Cederwall,
``Pure spinor superfields -- an overview,''
Springer Proc. Phys. {\bf 153}, 61-93 (2014).
[arXiv:1307.1762 [hep-th]].
%35 citations counted in INSPIRE as of 15 Dec 2022
}

\lref\tendsupertwistors{
N.~Berkovits,
``Ten-Dimensional Super-Twistors and Super-Yang-Mills,''
JHEP {\bf 04}, 067 (2010).
[arXiv:0910.1684 [hep-th]].
}

\lref\maxdiegoone{
D.~Garc\'\i{}a Sep\'ulveda and M.~Guillen,
``A pure spinor twistor description of the $D = 10$ superparticle,''
JHEP {\bf 08}, 130 (2020).
[arXiv:2006.06023 [hep-th]].
}

\lref\maxdiegotwo{
D.~G.~Sep\'ulveda and M.~Guillen,
``A Pure Spinor Twistor Description of Ambitwistor Strings,''
[arXiv:2006.06025 [hep-th]].
%2 citations counted in INSPIRE as of 16 Dec 2022
}

\lref\nmmax{
N.~Berkovits, M.~Guillen and L.~Mason,
``Supertwistor description of ambitwistor strings,''
JHEP {\bf 01}, 020 (2020).
[arXiv:1908.06899 [hep-th]].
}

\lref\maxmasoncasaliberkovits{
N.~Berkovits, E.~Casali, M.~Guillen and L.~Mason,
``Notes on the $D=11$ pure spinor superparticle,''
JHEP {\bf 08}, 178 (2019).
[arXiv:1905.03737 [hep-th]].
}

\lref\maxthesis{
M.~Guillen,
``Pure spinors and $D=11$ supergravity,''
[arXiv:2006.06014 [hep-th]].
}

\font\mbb=msbm10 
\newfam\bbb
\textfont\bbb=\mbb

\def\startcenter{%
  \par
  \begingroup
  \leftskip=0pt plus 1fil
  \rightskip=\leftskip
  \parindent=0pt
  \parfillskip=0pt
}
\def\stopcenter{%
  \par
  \endgroup
}

\listtoc
\writetoc
\filbreak

\newsec Introduction

\seclab\secone

\noindent  10D super-Yang-Mills and 11D supergravity at linearized level have been shown to be elegantly described in a manifestly super-Poincar\'e covariant manner by the quantization of the 10D and 11D superparticles, respectively, using pure spinor variables \refs{\psparticle, \ICTP, \pselevenparticle}. These objects were introduced for the first time in the context of the superstring in \superpoincarequantization, and then generalized to the study of supermembranes in \pssupermembrane. The full descriptions of maximally supersymmetric gauge theories, including the aforementioned theories, on pure spinor superspace were later discovered by Cederwall in a series of papers \refs{\neight, \nsix, \nfour, \pssugra, \psborninfeld,\pssreview}, by making use of the pure spinor superfield formalism. In this framework, the pure spinor actions take strikingly simple polynomial forms in a fundamental pure spinor superfield $\Psi$, and contain all the Batalin-Vilkovisky fields of the theories in study.

\medskip
The kinetic term of the pure spinor field theories presents the standard form ``$\Psi Q\Psi$'', where $Q$ is the ordinary non-minimal pure spinor BRST operator \NMPS. Consequently, the propagator of these theories is proportional to the so-called b-ghost, a negative ghost number composite operator satisfying the property $\{Q, b\} = {P^2 \over 2}$. This operator was first constructed in the pure spinor superstring, and shown to play a crucial role for computing several multiloop scattering amplitudes \refs{\NMPS,\mafraone,\mafratwo,\mafrathree}. Likewise, their properties have been shown to be substantial to design a covariant map between the pure spinor formalism and the conventional RNS setting \refs{\rnspsone,\rnspstwo}. 

\medskip
\noindent In a recent work \maxmaor, it has been shown that the pure spinor master action of 10D super-Yang-Mills in the gauge $b\Psi = Q\Omega$, for some $\Omega$, referred to as the generalized Siegel gauge, reproduces the same scattering amplitudes as those obtained from the open pure spinor superstring in the field-theory limit \stieberger. More interestingly, the kinematic numerators at any multiplicity were found to be proportional to nested b-ghost expressions, and to match the multiparticle superfields constructed in \mafraoli\ up to generalized gauge transformations and BRST-exact terms. These computations were possible to be methodically carried out due to the existence of simpler alternative expressions for the 10D b-ghost \refs{\dynamical,\xiyin,\chiralmax}. Such expressions make use of negative ghost number operators, referred to as physical operators, satisfying a set of defining relations resembling the 10D super-Yang-Mills equations of motion at linearized order. Remarkably, these very same operators were ingeniously used to show that the Siegel gauge condition $b\Psi = 0$, implies a Poisson algebra structure for kinematic numerators, elegantly thus realizing the kinematic algebra of the Bern-Carrasco-Johansson (BCJ) duality \bcjone\ from an action principle viewpoint\foot{As discussed in \maxmaor, this statement is sensitive to the actual ability of computing pure spinor correlators in a certain regularization scheme. Hence, higher-loop generalizations of this kinematic algebra might be subtle due to the highly non-local behavior of pure spinor kinematic numerators.}.

\medskip 
In this work, we introduce the 11D analogues of the physical operators above mentioned, and provide a novel compact formula for the 11D b-ghost, introduced for the first time in \maximalloopcederwall, which will make computations involving the b-ghost more tractable and efficient. To illustrate this, we show that $\{Q , b\} = {P^{2} \over 2}$ and $\{b,b\} = Q\Omega$, in a straightforward and elegant way, as a consequence of the simple properties satisfied by the physical operators. In addition, we use our new formula to construct a ghost number two vertex operator via a standard descent relation involving the ghost number three vertex operator. Up to BRST-exact terms, the operator thus obtained is shown to match that introduced in \maxnotesworldline\ using the Y-formalism \OdaTonin\ in 11D. Furthermore, we find that the ghost number three operator satisfies the generalized Siegel gauge condition after letting the b-ghost act on it as a second-order differential operator. Finally, we apply the perturbiner method \perturbiner\ to the pure spinor description of 11D supergravity, and by making use of our new formula for the b-ghost, we readily solve the two-particle superfield equation of motion.

\medskip
This paper is organized as follows. In section 2, we review the non-minimal pure spinor construction of the 11D superparticle, and discuss the formulae found for the b-ghost in \maximalloopcederwall\ and \elevendsimplifiedb. In section 3, we introduce the 11D physical operators, and compute their actions on the ghost number three vertex operator. We then write down a compact formula for the 11D b-ghost in terms of the physical operators and, after full expansion, it is shown to coincide with the original proposal in \maximalloopcederwall. In section 4, we give some applications showing how our new formula for the b-ghost considerably simplifies computations relevant to scattering processes in 11D supergravity. We close with discussions and future directions in section 5. Appendix A is devoted to a short review of the superspace equations of motion of linearized 11D supergravity, and Appendix B spells out the 11D pure spinor projector used in this work.

\newsec 11D Non-minimal pure spinor superparticle 

\seclab\secthree

\noindent The 11D pure spinor superparticle action in flat space is defined by \refs{\pselevenparticle,\pssupermembrane}
\eqnn \elevendpsaction
$$ \eqalignno{
S &= \int d\tau [P^a \partial_{\tau}X_a + p_{\alpha}\partial_{\tau}\theta^{\alpha} + w_{\alpha}\partial_{\tau}\lambda^{\alpha} - \half P^2] & \elevendpsaction
}
$$
We will use letters from the beginning of the Greek/Latin alphabet to denote spinor/vector $SO(1,10)$ indices. The variables $(X^a, \theta^{\alpha})$ are the usual 11D superspace coordinates, and $(P_a, p_{\alpha})$ are their respective conjugate momenta. The bosonic spinor $\lambda^{\alpha}$ satisfies the 11D pure spinor constraint, i.e. $\lambda\gamma^a \lambda = 0$. Its respective conjugate momentum $w_{\alpha}$ is thus only defined up to the gauge transformation $\delta w_{\alpha} = (\gamma^{a}\lambda)_{\alpha}\sigma_a$, for any vector $\sigma_{a}$. Due to their wrong statistics, they will be called ghosts and assigned to carry ghost numbers 1 and -1, respectively. The 11D gamma matrices will be represented by $(\gamma^{a})_{\alpha\beta}$, $(\gamma^{a})^{\alpha\beta}$, and they satisfy the Clifford algebra: $(\gamma^{a})_{\alpha\beta}(\gamma^{b})^{\beta\delta} + (\gamma^{b})_{\alpha\beta}(\gamma^{a})^{\beta\delta} = 2\eta^{ab}\delta_{\alpha}^{\delta}$. We will raise and lower spinor indices by using the antisymmetric charge conjugation matrix $C_{\alpha\beta}$ and its inverse $C^{\alpha\beta}$, which obey the relation $C_{\alpha\beta}C^{\beta\delta} = \delta_{\alpha}^{\delta}$, so that $(\gamma^{a})^{\alpha\beta} = C^{\alpha\epsilon}C^{\beta\delta}(\gamma^{a})_{\epsilon\delta}$, etc.

\medskip
\noindent The Hilbert space is described by the BRST-cohomology of the operator $Q_{0} = \lambda^{\alpha} d_{\alpha}$, where $d_{\alpha}$ is the Brink-Schwarz fermionic constraint \brinkschwarz\ defined as
\eqnn \elevendpsbrst
$$ \eqalignno{
d_{\alpha} &= p_{\alpha} - \half (\gamma^a \theta)_{\alpha}P_{a} & \elevendpsbrst
}
$$
Such a cohomology can be shown to be non-trivial up to ghost number 7, describing the 11D supergravity states in its Batalin-Vilkovisky formulation. Concretely, the ghost number 0, 1, 2 and 3 sectors respectively host the gauge symmetry ghost-for-ghost-for-ghost; the gauge symmetry ghost-for-ghost; the supersymmetry, diffeomorphism and gauge symmetry ghosts; and the 11D supergravity physical fields. The higher ghost number sectors form a mirror cohomology of those above described, and reproduce the 11D supergravity antifields. In order to illustrate this, let us analyze the cohomology at ghost number three, $U^{(3)} = \Psi = \lambda^{\alpha}\lambda^{\beta}\lambda^{\delta}A_{\alpha\beta\delta}$. The physical state conditions then imply that
\eqnn \closed
\eqnn \exact
$$ \eqalignno{
Q_{0}\Psi &= 0  \rightarrow D_{(\alpha}A_{\beta\delta\epsilon)} = (\gamma^{a})_{(\alpha\beta}A_{a\delta\epsilon)} & \closed \cr
\delta \Psi &= Q_{0}\Lambda \rightarrow \delta A_{\alpha\beta\delta} = D_{(\alpha}\Lambda_{\beta\delta)} & \exact
}
$$
where $\Lambda = \lambda^{\alpha}\lambda^{\beta}\Lambda_{\alpha\beta}$, and $\Lambda_{\alpha\beta}$ is any superfield. These equations match the linearized equations of motion of 11D supergravity in superspace \brinkhowe, we thus identify $A_{\alpha\beta\delta} = C_{\alpha\beta\delta}$, where $C_{\alpha\beta\delta}$ is the linearized version of the lowest-dimensional component of the 11D supergravity super-3-form. In a particular gauge, one can show that $\Psi$ has the following $\theta$-expansion:
\eqnn \psithetaexpansion
$$ \eqalignno{
\Psi =& (\lambda\gamma^{a}\theta)(\lambda\gamma^{b}\theta)(\lambda\gamma^{c}\theta)C_{abc} + (\lambda\gamma^{ab}\theta)(\lambda\gamma_{b}\theta)(\lambda\gamma^{c}\theta)h_{ac}  + (\lambda\gamma^{a}\theta)(\lambda\gamma^{b}\theta)(\lambda\gamma^{c}\theta)(\theta\gamma_{bc}\psi_{a})\cr
& - (\lambda\gamma^{a}\theta)(\lambda\gamma^{bc}\theta)(\lambda\gamma_{b}\theta)(\theta\gamma_{c}\psi_{a}) + O(\theta^5) & \psithetaexpansion
}
$$
with $C_{abc}$, $h_{ab}$, $\psi^a_{\alpha}$ being respectively the 3-form, graviton and gravitino of 11D supergravity. Indeed, they can be shown to satify the linearized equations of motion 
\eqnn \equationsofmotion
$$\eqalignno{
\partial^{d}\partial_{[d}C_{abc]} &= 0 \ , \ \ \lform h_{bc} - 2 \partial^{a}\partial_{(b}h_{c)a} + \partial_{b}\partial_{c}(\eta^{ad}h_{ad}) = 0 \ , \ \ (\gamma^{abc})_{\alpha\beta}\partial_{b}\psi_{c}^{\beta} = 0  & \equationsofmotion
}
$$
and gauge transformations
\eqnn \gaugetransformations
$$\eqalignno{
\delta C_{abc} &= \partial_{a}B_{bc} \ , \ \ \delta h_{ab} = \partial_{(a}t_{b)} \ , \ \ \delta \psi_{a}^{\alpha} = \partial_{a}\kappa^{\beta} & \gaugetransformations
}
$$
where $B_{ab}$, $t_{b}$ and $\kappa^{\beta}$ are arbitrary gauge parameters.

\medskip
In order to define negative ghost number pure spinor operators, one needs to introduce the so-called non-minimal pure spinor variables \NMPS. These ones consist of two pairs of conjugate variables $(\bar{\lambda}_{\alpha},\bar{w}^{\beta})$, $(r_{\alpha}, s^{\beta})$, where $\bar{\lambda}_{\alpha}$ is a ghost number -1 pure spinor variable satisfying $\bar{\lambda}\gamma^{a}\bar{\lambda} = 0$, and $r_{\alpha}$ is a ghost number 0 fermionic spinor constrained via $\bar{\lambda}\gamma^{a}r = 0$. The 11D non-minimal pure spinor superparticle is then defined by the action \elevendsimplifiedb
\eqnn \nonminimalaction
$$\eqalignno{
S &= \int d\tau [P^a \partial_{\tau}X_a + p_{\alpha}\partial_{\tau}\theta^{\alpha} + w_{\alpha}\partial_{\tau}\lambda^{\alpha} + \bar{w}^{\alpha}\partial_{\tau}\bar{\lambda}_{\alpha} + s^{\alpha}\partial_{\tau}r_{\alpha} - \half P^2] & \nonminimalaction
}
$$
together with the BRST operator
\eqnn \nonminimalbrstoperator
$$\eqalignno{
Q &= Q_{0} + s & \nonminimalbrstoperator
}
$$
where $s = r_{\alpha}\bar{w}^{\alpha}$. Using the quartet mechanism \quartet, one can show that the cohomology of $Q$ will be independent of the non-minimal variables, therefore matching that of $Q_{0}$.

\subsec The b-ghost

\subseclab\secthreeone

As in 10D, it is possible to construct the so-called b-ghost, a ghost number -1 operator, obeying $\{Q, b\} = \half P^2$. This object was originally constructed in \maximalloopcederwall, and shown to take the complicated form
%\eqnn \bghost
%$$\eqalignno{
%b =& {1\over 2\eta}(\bar{\lambda}\gamma^{ab}\bar{\lambda})(\lambda\gamma^{ab}\gamma^{c}d)P_{c} + {1\over\eta^2}L^{(1)}_{ab,cd}\bigg[(\lambda\gamma^{a}d)(\lambda\gamma^{bcd}d) + 2(\lambda\gamma^{abcef}\lambda)N^{d}{}_{e}P_{f}\cr
%& + {2 \over 3}(\delta^{b}_{e}\delta^{d}_{f} - \eta^{bd}\eta_{ef})(\lambda\gamma^{aecgh}\lambda)N_{gh}P^{f}\bigg] - {1\over 3 \eta^3} L^{(2)}_{ab,cd,ef}\bigg[(\lambda\gamma^{abcgh}\lambda)(\lambda\gamma^{def}d)N_{gh} \cr 
%& - 12[(\lambda\gamma^{abceg}\lambda)\eta^{fh} - {2\over 3}\eta^{f[a}(\lambda\gamma^{bce]gh}\lambda)](\lambda\gamma^{d}d)N_{gh}\bigg]\cr
%& + {4\over 3 \eta^4}L^{(3)}_{ab,cd,ef,gh}(\lambda\gamma^{abcij}\lambda)\bigg[(\lambda\gamma^{defgk}\lambda)\eta^{hl} - {2\over 3}\eta^{h[d}(\lambda\gamma^{efgk]l}\lambda)\{N_{ij}, N_{kl}\}\bigg] & \bghost }
%$$
\eqnn \bghost
$$\eqalignno{
b =& {1\over 2\eta}(\bar{\lambda}\gamma^{ab}\bar{\lambda})(\lambda\gamma^{ab}\gamma^{c}d)P_{c} + {2\over\eta^2}L^{(1)}_{ab,cd}\bigg[(\lambda\gamma^{a}d)(\lambda\gamma^{bcd}d) + 2(\lambda\gamma^{abcef}\lambda)N^{d}{}_{e}P_{f}\cr
& + {2 \over 3}(\delta^{b}_{e}\delta^{d}_{f} - \eta^{bd}\eta_{ef})(\lambda\gamma^{aecgh}\lambda)N_{gh}P^{f}\bigg] - {4\over 3 \eta^3} L^{(2)}_{ab,cd,ef}\bigg[(\lambda\gamma^{abcgh}\lambda)(\lambda\gamma^{def}d)N_{gh} \cr 
& - 12[(\lambda\gamma^{abceg}\lambda)\eta^{fh} - {2\over 3}\eta^{f[a}(\lambda\gamma^{bce]gh}\lambda)](\lambda\gamma^{d}d)N_{gh}\bigg]\cr
& + {8\over 3 \eta^4}L^{(3)}_{ab,cd,ef,gh}(\lambda\gamma^{abcij}\lambda)\bigg[(\lambda\gamma^{defgk}\lambda)\eta^{hl} - {8\over 3}\eta^{h[d}(\lambda\gamma^{efgk]l}\lambda)\bigg]\{N_{ij}, N_{kl}\} & \bghost
} 
$$
where $\eta = (\bar{\lambda}\gamma^{ab}\bar{\lambda})(\lambda\gamma_{ab}\lambda)$, $N_{ab} = \half(\lambda\gamma^{ab}w)$ is the usual ghost Lorentz current, and $L^{(n)}_{a_0 b_0, a_1 b_1, \ldots, a_1 b_1} = (\bar{\lambda}\gamma_{[[a_0 b_0}\bar{\lambda})(\bar{\lambda}\gamma_{a_1 b_1}r)\ldots (\bar{\lambda}\gamma_{a_n b_n]]}r)$, with $[[\ ]]$ denoting antisymmetrization between each pair of indices. Remarkably, this operator was simplified in \elevendsimplifiedb\ to the simpler expression
\eqnn \simplifiedbghost
$$\eqalignno{
b &= P^{a}\bar{\Sigma}_{a} - {4\over \eta}(\bar{\lambda}\gamma^{ab}r)(\lambda\gamma_{a}{}^{c}\lambda)\bar\Sigma_{c}\bar{\Sigma}_{b} - {2\over \eta}(\bar{\lambda}r)(\lambda\gamma^{ab}\lambda)\bar{\Sigma}_{a}\bar{\Sigma}_{b} & \simplifiedbghost
}
$$
where the fermionic vector $\bar{\Sigma}^{i}$, defined by
\eqnn \sigmaa
$$ \eqalignno{
\bar{\Sigma}^{i} =& {1 \over 2\eta} (\bar{\lambda} \gamma^{ab}\bar{\lambda})(\lambda\gamma^{ab}\gamma^{i}d) + {4\over \eta^2} L^{(1)}_{ab,cd}(\lambda\gamma^{abcei}\lambda)N^{d}{}_{e} + {4 \over 3 \eta^2}L^{(1)}_{ab,c}{}^{i}(\lambda\gamma^{abcde}\lambda)N_{de}\cr 
& - {4\over 3 \eta^2}L^{(1)}_{ad,c}{}^{d}(\lambda\gamma^{aicde}\lambda)N_{de} & \sigmaa
}
$$
obeys $(\bar{\lambda}\gamma^{ab}\bar{\lambda})\bar{\Sigma}_{b} = 0$, and
\eqnn \qsigmaa
$$\eqalignno{
\{Q, \bar{\Sigma}^{a}\} =& {P^{a} \over 2} + {1\over \eta}[(\bar{\lambda}\gamma^{cb}\bar{\lambda})(\lambda\gamma_{ba}\lambda) - (\bar{\lambda}\gamma^{ab}\bar{\lambda})(\lambda\gamma_{bc}\lambda)]P^{c} - {2\over \eta}(\bar{\lambda}\gamma^{ba}r)(\lambda\gamma_{b}{}^{c}\lambda)\bar{\Sigma}_{c}\cr
& - {4\over \eta}(\bar{\lambda}\gamma^{bc}r)(\lambda\gamma_{b}{}^{a}\lambda)\bar{\Sigma}_{c} + {2\over \eta}(\bar{\lambda}r)(\lambda\gamma^{ab}\lambda)\bar{\Sigma}_{b} - {2\over \eta^2}(\bar{\lambda}\gamma^{cd}r)(\lambda\gamma_{cd}\lambda)(\bar{\lambda}\gamma^{ab}\bar{\lambda})(\lambda\gamma_{be}\lambda)\bar{\Sigma}^{e}\cr & & \qsigmaa
}
$$
Using the identity \qsigmaa, it was shown in \elevendsimplifiedb, the simplified expression \simplifiedbghost\ indeed satisfies the property $\{Q ,b\} = \half P^2$, and it is nilpotent up to BRST-exact terms. 

\newsec 11D Physical operators 

\seclab\secfour

In this section we introduce the 11D analogues of the operators studied in \psborninfeld\ in the 10D case. These will be proven to be essential for a new formulation of the b-ghost exhibiting its close relation to 11D supergravity.

\subsec Physical operators

\subseclab\secfourone

The 11D physical operators will be defined as follows
\eqnn \drchatalpha
\eqnn \drchata
\eqnn \drphihata
\eqnn \drphihatalpha
$$ \eqalignno{
[Q, {\bf C}_{\alpha}] &= -{1\over 3}d_{\alpha} - (\gamma^a \lambda)_{\alpha}{\bf C}_{a}  & \drchatalpha \cr
\{Q, {\bf C}_{a}\} &= {1\over 3}P_{a} + (\lambda\gamma^{ab}\lambda){\bf \Phi}_{b} 
 & \drchata \cr
[Q, {\bf \Phi}^{a}] &= (\lambda\gamma^a {\bf \Phi}) & \drphihata \cr
[Q, {\bf \Phi}^{\alpha}] &= {1\over 4}(\lambda\gamma^{ab})^{\alpha}{\bf \Omega}_{ab} & \drphihatalpha \cr
\vdots
}
$$
These relations follow immediately from the linearized 11D supergravity equations of motion (see Appendix A for a short review). The elipsis below \drphihatalpha\ represent additional equations which will not be relevant for our purposes. The system of equations above displayed is solved by
\eqnn \chatalpha
\eqnn \chata
\eqnn \phihata
\eqnn \phihatalpha
$$
\eqalignno{
{\bf C}_{\alpha} &= {1\over 3}K_{\alpha}{}^{\beta}w_{\beta}  & \chatalpha\cr
{\bf C}^{a} &= {1 \over \eta} (\lambda\gamma^{abc})^{\alpha}(\bar{\lambda}\gamma_{bc}\bar{\lambda})\bigg[{1 \over 3}d_{\alpha} + [Q, {\bf C}_{\alpha}]\bigg] & \chata \cr 
{\bf \Phi}^{a} &= {2 \over \eta} (\bar{\lambda}\gamma^{ab}\bar{\lambda})\bigg[{1\over 3}P_{b} - \{Q, {\bf C}_{b}\}\bigg] & \phihata \cr
{\bf \Phi}^{\alpha} &= -{2\over \eta}(\gamma^{abc}\lambda)^{\alpha}(\bar{\lambda}\gamma_{bc}r){\bf \Phi}_{a} & \phihatalpha
}
$$
where $K_{\alpha}{}^{\beta}$ is an 11D pure spinor projector defined as
\eqnn \psprojectormine
$$
\eqalignno{
K_{\alpha}{}^{\beta} =& - {1\over 6\eta}(\lambda\gamma^{ab})^{\beta}(\bar{\lambda}\gamma^{cd}\bar{\lambda})(\lambda\gamma_{abcd})_{\alpha} - {4\over 3\eta}(\lambda\gamma^{ab})^{\beta}(\lambda\gamma_{b}{}^{d})_{\alpha}(\bar{\lambda}\gamma_{ad}\bar{\lambda}) - {2\over 3\eta}(\lambda\gamma^{cd})^{\beta}\lambda_{\alpha}(\bar{\lambda}\gamma_{cd}\bar{\lambda}) & \cr
& + {1\over 3\eta}\lambda^{\beta}(\lambda\gamma^{cd})_{\alpha}(\bar{\lambda}\gamma_{cd}\bar{\lambda}) & \psprojectormine
}
$$
and the operators are constrained to satisfy
\eqnn \constraintsforop
$$
\eqalignno{
\xi_{a}^{\alpha}{\bf C}_{\alpha} &= 0, \ \ \ \  \ (\bar{\lambda}\gamma^{ab}\bar{\lambda}){\bf C}_{a} = 0, \ \ \ \ \ (\bar{\lambda}\gamma^{a})_{\alpha}{\bf \Phi}_{a} = 0, \ \ \ \ \ R_{\alpha}{}^{\beta}{\bf \Phi}^{\alpha} = 0 & \constraintsforop
}
$$
with $\xi_{a}^{\beta}$ and $R_{\alpha}{}^{\beta}$ taking the explicit forms
\eqnn \xiaalpha
\eqnn \ralphabeta
$$
\eqalignno{
\xi_{a}^{\beta} =& \half (\gamma_{abc})^{\beta\delta}\lambda_{\delta}(\bar{\lambda}\gamma^{bc}\bar{\lambda}) &  \xiaalpha\cr
R_{\alpha}{}^{\beta}=& \bigg[-{1\over 2}(\lambda\gamma^{b})_{\alpha}(\lambda\gamma^{c})^{\beta} - {1\over 4}(\lambda\gamma^{bk}\lambda)(\gamma^{c}\gamma^{k})_{\alpha}{}^{\beta} + {1\over 2}(\lambda\gamma^{bk})_{\alpha}(\lambda\gamma^{ck})^{\beta} - {1\over 2}(\lambda\gamma^{bc})_{\alpha}\lambda^{\beta}\bigg](\bar{\lambda}\gamma_{bc}\bar{\lambda}) &  \cr 
&  &\ralphabeta
}
$$
%The first term in $R_alpha^beta$ has the unconventional structure $\lambda^{\delta}(\gamma^{b})_{\delta\alpha}\lambda^{\nu}(\gamma^{c})_{\nu}{}^{\beta}$
These objects were previously defined in \maxnotesworldline\ where they were shown to play an important role in the construction of a ghost number -2 operator mapping the cohomology of the ghost number three vertex operator into that of the ghost number one vertex operator. They obey the useful relation $(\lambda\gamma^{a})_{\alpha}\xi_{a}^{\beta} = \half \delta_{\alpha}^{\beta}\eta + R_{\alpha}{}^{\beta}$.

\medskip
\noindent The projector of eqn. \psprojectormine\ satisfies the desired properties
\eqnn \propertiesofk
$$
\eqalignno{
\lambda^{\alpha}K_{\alpha}{}^{\beta} &= \lambda^{\beta}, \ \ \ \ \ (\lambda\gamma^{ab})^{\alpha}K_{\alpha}{}^{\beta} = (\lambda\gamma^{ab})^{\beta}, \ \ \ \ \ (\lambda\gamma^{a})_{\beta}K_{\alpha}{}^{\beta} = 0, \ \ \ \ \ K_{\alpha}{}^{\beta}K_{\beta}{}^{\delta} = K_{\alpha}{}^{\delta} & \cr
& & \propertiesofk
}
$$
and its trace can be shown to match the dimension of the 11D pure spinor space, that is $K_{\alpha}{}^{\alpha} = 23$. This statement can easily be proven by rewriting $K_{\alpha}{}^{\beta}$ in the more compact form
\eqnn \psprojectorminecompact
$$
\eqalignno{
K_{\alpha}{}^{\beta} &= \delta_{\alpha}^{\beta} + {1\over \eta}(\lambda\gamma^{abc})^{\beta}(\bar{\lambda}\gamma_{bc}\bar{\lambda})(\lambda\gamma_{a})_{\alpha} & \psprojectorminecompact
}
$$
A demonstration of the equivalence between eqns. \psprojectormine\ and \psprojectorminecompact\ is provided in Appendix B.

\medskip
Explicitly, the physical operators read
\eqnn \chatalphaexp
\eqnn \chataexp
\eqnn \phihataexp
\eqnn \phihatalphaexp
$$
\eqalignno{
{\bf C}_{\alpha} =& {w_{\alpha} \over 3} + {1\over 3\eta}(\lambda\gamma^{abc}w)(\bar{\lambda}\gamma_{bc}\bar{\lambda})(\lambda\gamma_{a})_{\alpha} & \chatalphaexp \cr
{\bf C}_{a} =& {1\over 3\eta}(\bar{\lambda}\gamma^{bc}\bar{\lambda})(\lambda\gamma_{abc}d) - {2\over 3\eta}(\bar{\lambda}\gamma^{bc}r)(\lambda\gamma_{abc}w) + {2\over 3\eta^2}\phi (\bar{\lambda}\gamma^{bc}\bar{\lambda})(\lambda\gamma_{abc}w) & \cr
& +{4\over 3 \eta^{2}}(\lambda\gamma_{ac}\lambda)(\bar{\lambda}\gamma^{bc}\bar{\lambda})(\bar{\lambda}\gamma^{de}r)(\lambda\gamma_{bde}w)  & \chataexp \cr
{\bf \Phi}^{a} =& {2 \over 3}\bigg[{1 \over \eta}(\bar{\lambda}\gamma^{ab}\bar{\lambda})P_{b} - {2 \over \eta^{2}}(\bar{\lambda}\gamma^{ab}\bar{\lambda})(\bar{\lambda}\gamma^{cd}r)(\lambda\gamma_{bcd}d) + \{s, {2\over \eta^2}(\bar{\lambda}\gamma^{ab}\bar{\lambda})(\bar{\lambda}\gamma^{cd}r)\}(\lambda\gamma_{bcd}w) & \cr
& - {8 \over \eta^3}(\lambda\gamma^{a}\xi_{b})(\bar{\lambda}\gamma^{cb}r)(\bar{\lambda}\gamma^{de}r)(\lambda\gamma_{cde}w)\bigg] & \phihataexp \cr
{\bf \Phi}^{\alpha} =& {8 \over 3}\xi_{a}^{\alpha}\bigg[{1\over \eta^{2}}(\bar{\lambda}\gamma^{ab}r)P_{b} - {4 \over \eta^{4}}(\bar{\lambda}\gamma^{ab}r)(\lambda\gamma_{cb}\lambda)(\bar{\lambda}\gamma^{cd}\bar{\lambda})(\bar{\lambda}\gamma^{ef}r)(\lambda\gamma_{def}d) & \cr
& - \bigg({8 \over \eta^{4}} (\bar{\lambda}\gamma^{ab}r)(\lambda\gamma_{cb}\lambda)(\bar{\lambda}\gamma^{cd}r)(\bar{\lambda}\gamma^{ef}r) - {16 \over \eta^{5}}(\bar{\lambda}\gamma^{ab}r)\phi (\lambda\gamma_{cb}\lambda)(\bar{\lambda}\gamma^{cd}\bar{\lambda})(\bar{\lambda}\gamma^{ef}r)\bigg)(\lambda\gamma_{def}w)\bigg] & \cr
&  & \phihatalphaexp 
}
$$
where $\phi = (\lambda\gamma^{ab}\lambda)(\bar{\lambda}\gamma_{ab}r)$. These relations can be rewritten in a manifestly gauge invariant form, as follows
\eqnn \chatalphaexpgi
\eqnn \chataexpgi
\eqnn \phihataexpgi
$$
\eqalignno{
{\bf C}_{\alpha} =& - {1\over 9\eta}N^{ab}(\bar{\lambda}\gamma^{cd}\bar{\lambda})(\lambda\gamma_{abcd})_{\alpha} - {8\over 9\eta}N^{ab}(\lambda\gamma_{b}{}^{d})_{\alpha}(\bar{\lambda}\gamma_{ad}\bar{\lambda}) - {4\over 9\eta}N^{cd}\lambda_{\alpha}(\bar{\lambda}\gamma_{cd}\bar{\lambda}) & \cr
& + {2\over 9\eta}J(\lambda\gamma^{cd})_{\alpha}(\bar{\lambda}\gamma_{cd}\bar{\lambda}) & \chatalphaexpgi \cr
{\bf C}_{a} =& {1\over 3\eta}(\lambda\gamma_{abc}d)(\bar{\lambda}\gamma^{bc}\bar{\lambda}) + {8\over 3\eta^{2}}(\bar{\lambda}\gamma^{bc}\bar{\lambda})(\bar{\lambda}\gamma^{de}r)(\lambda\gamma_{bcdfa}\lambda)N_{e}{}^{f} + {8 \over 9\eta^2} (\bar{\lambda}\gamma^{bc}\bar{\lambda})(\bar{\lambda}r)(\lambda\gamma_{abcde}\lambda)N^{de} &  \cr
& + {4 \over 9\eta^2}(\bar{\lambda}\gamma_{bc}\bar{\lambda})(\bar{\lambda}\gamma_{da}r)(\lambda\gamma^{bcdef}\lambda)N_{ef} & \chataexpgi \cr
{\bf \Phi}^{a} =& {2 \over 3}\bigg[{1\over \eta}(\bar{\lambda}\gamma^{ab}\bar{\lambda})P_{b} - {2\over \eta^{2}}(\bar{\lambda}\gamma^{ab}\bar{\lambda})(\bar{\lambda}\gamma^{cd}r)(\lambda\gamma_{bcd}d) - {16\over \eta^3}(\bar{\lambda}\gamma^{ab}\bar{\lambda})(\bar{\lambda}\gamma^{cd}r)(\bar{\lambda}\gamma^{ef}r)(\lambda\gamma_{bcdeg}\lambda)N_{f}{}^{g} & \cr
& - {8 \over \eta^{3}}(\bar{\lambda}\gamma^{ab}\bar{\lambda})(\bar{\lambda}\gamma^{cd}r)(\bar{\lambda}r)(\lambda\gamma_{bcdef}\lambda)N^{ef}\bigg] & \phihataexpgi \cr
{\bf \Phi}^{\alpha} =& {8 \over 3}\xi_{a}^{\alpha}\bigg[{1\over \eta^2}(\bar{\lambda}\gamma^{ab}r)P_{b} - {2\over \eta^{3}}(\bar{\lambda}\gamma^{ab}r)(\bar{\lambda}\gamma^{cd}r)(\lambda\gamma_{bcd}d)- {8 \over \eta^{4}}(\bar{\lambda}\gamma^{ab}r)(\bar{\lambda}\gamma^{cd}r)(\bar{\lambda}r)(\lambda\gamma_{bcdef}\lambda)N^{ef} & \cr
&  - {16\over \eta^{4}}(\bar{\lambda}\gamma^{ab}r)(\bar{\lambda}\gamma^{cd}r)(\bar{\lambda}\gamma^{ef}r)(\lambda\gamma_{bcdeg}\lambda)N_{f}{}^{g} \bigg] &
}
$$
where $J = \lambda^{\alpha} w_{\alpha}$.

\medskip
\noindent Now it is easy to calculate the action of the 11D physical operators on $\Psi$. Let us start with ${\bf C}_{\alpha}$. The formula \psprojectorminecompact\ immediately implies that
\eqnn \chatalphaonpsi
$$
\eqalignno{
{\bf C}_{\alpha}\Psi &= C_{\alpha} + (\lambda\gamma^{a})_{\alpha}\rho_{a} & \chatalphaonpsi
}
$$
where $C_{\alpha} = \lambda^{\beta}\lambda^{\delta}C_{\alpha\beta\delta}$, and $\rho^{a} = {1\over \eta}(\lambda\gamma^{abc})^{\alpha}C_{\alpha}(\bar{\lambda}\gamma_{bc}\bar{\lambda})$. Using eqn. \chatalphaonpsi, one can compute the action of ${\bf C}^{a}$ on $\Psi$. Indeed, one finds that
\eqnn \chataonpsi
$$
\eqalignno{
{\bf C}_{a}\Psi &= C_{a} + (\lambda\gamma_{ac}\lambda)s^{c} - Q\rho_{a} & \chataonpsi
}
$$
where $C_{a} = \lambda^{\beta}\lambda^{\delta}C_{a\beta\delta}$, and $s^{b} = -{2\over \eta}(\bar{\lambda}\gamma^{bc}\bar{\lambda})(C_{c}-Q\rho_{c})$. Similarly, the use of eqn. \chataonpsi\ allows one to show that
\eqnn \phihataonpsi
$$
\eqalignno{
{\bf \Phi}^{a}\Psi &= \Phi^{a} + (\lambda\gamma^{a}\kappa) + Qs^a & \phihataonpsi 
}
$$
where $\Phi^{a} = \lambda^{\alpha}h_{\alpha}{}^{a}$, and $\kappa^{\alpha} = -2\xi^{\alpha}_{a}(\Phi^{a} + Qs^{a})$. Finally, eqn. \phihataonpsi\ implies that the action of ${\bf \Phi}^{\alpha}$ on $\Psi$ is given by
\eqnn \phihatalphaonpsi
$$
\eqalignno{
{\bf \Phi}^{\alpha}\Psi &= \Phi^{\alpha} + (\lambda\gamma^{ab})^{\alpha}f_{ab} + \lambda^{\alpha}f  + Q\kappa^{\beta} & \phihatalphaonpsi
}
$$
where $\Phi^{\alpha} = \lambda^{\beta}h_{\beta}{}^{\alpha}$, $f_{ab} = {2\over 3\eta}(\bar{\lambda}\gamma_{ab}\bar{\lambda})\lambda_{\delta} \tau^{\delta} +  {4\over 3\eta}(\bar{\lambda}\gamma_{k[a}\bar{\lambda})(\lambda\gamma^{k}{}_{b]})_{\alpha}\tau^{\alpha} + {1 \over 6\eta}(\bar{\lambda}\gamma^{cd}\bar{\lambda})(\lambda\gamma_{cdab})_{\alpha}\tau^{\alpha}$, $f = -{1\over 3\eta} (\bar{\lambda}\gamma_{ab}\bar{\lambda})(\lambda\gamma^{ab})_{\delta}\tau^{\delta}$, $\tau^{\alpha} = \Phi^{\alpha} + Q\kappa^{\alpha}$, and we used the alternative expression for $R_{\alpha}{}^{\beta}$
\eqnn \alternativeralphabeta
$$
\eqalignno{
R_{\alpha}{}^{\beta} &= \bigg[{1 \over 12}(\lambda\gamma^{abcd})_{\alpha}(\lambda\gamma_{ab})^{\beta} + {2\over 3}(\lambda\gamma^{kd})_{\alpha}(\lambda\gamma^{c}{}_{k})^{\beta}  + {1\over 3}\lambda_{\alpha}(\lambda\gamma^{cd})^{\beta}  - {1\over 6}(\lambda\gamma^{cd})_{\alpha}\lambda^{\beta}\bigg](\bar{\lambda}\gamma_{cd}\bar{\lambda}) & \cr
%R_{\alpha}{}^{\beta} &= \bigg[{2\over 3}\lambda_{\alpha}(\lambda\gamma^{cd})^{\beta} + {4\over 3}(\lambda\gamma^{kd})_{\alpha}(\lambda\gamma^{ck})^{\beta} - {5\over 6}(\lambda\gamma^{cd})_{\alpha}\lambda^{\beta} - {1\over 12}(\lambda\gamma^{cdab})_{\alpha}(\lambda\gamma^{ab})^{\beta}\bigg](\bar{\lambda}\gamma_{cd}\bar{\lambda}) & \cr
%R_{\alpha}{}^{\beta} &= \bigg[{1\over 3}\lambda_{\alpha}(\lambda\gamma^{cd})^{\beta} + {7\over 6}(\lambda\gamma^{ck})_{\alpha}(\lambda\gamma^{dk})^{\beta} - {2\over 3}(\lambda\gamma^{cd})_{\alpha}\lambda^{\beta} + {1\over 12}(\lambda\gamma^{cdab)_{\alpha}(\lambda\gamma^{ab})^{\beta}\bigg](\bar{\lambda}\gamma_{cd}\bar{\lambda}) & \cr
& & \alternativeralphabeta
}
$$

\subsec A simple expression for the b-ghost

\subseclab \secfourtwo

\noindent
The physical operators recently studied allow us to write the following alternative expression for the 11D b-ghost:
\eqnn \physicalbghost
$$ \eqalignno{
b &= {3 \over 2}P^a {\bf C}_{a} + {3 \over 2}(\lambda\gamma^{a}d){\bf \Phi}_{a} - {3 \over 2}(\lambda\gamma^{a}w)(\lambda\gamma_{a}{\bf \Phi}) & \physicalbghost
}
$$
or in a gauge invariant form
\eqnn \physicalbghostgi
$$ \eqalignno{
b &= {3 \over 2}P^a {\bf C}_{a} + {3 \over 2}(\lambda\gamma^{a}d){\bf \Phi}_{a} - {1 \over 2}N^{ab}(\lambda\gamma_{ab}{\bf \Phi}) & \physicalbghostgi
}
$$
It is easy to show that $\{Q, b\} = {P^2 \over 2}$. Indeed, the use of the defining properties \drchatalpha-\drphihatalpha\ imply that
\eqnn \showingqbp
$$
\eqalignno{
\{Q, b\} =& {1\over 2} P^2 + {3\over 2}(\lambda\gamma^{ab}\lambda)P_{a}{\bf \Phi}_{b} + {3\over 2}(\lambda\gamma^{ab}\lambda)P_{b}{\bf \Phi}_{a} - {3 \over 2}(\lambda\gamma^{a}d)(\lambda\gamma_{a}{\bf \Phi}) + {3 \over 2}(\lambda\gamma^{a}d)(\lambda\gamma_{a}{\bf \Phi}) & \cr
 =& {1\over 2} P^2 & \showingqbp
 }
$$
One can also check that $b$ is nilpotent up to BRST-exact terms. To see this, it is enough to show that $\{b, b\}$ does not contain any term independent of $r_{\alpha}$ \elevendsimplifiedb. This easily follows from the explicit relations \chatalphaexp-\phihatalphaexp, and the convenient rewriting
\eqnn \bbbrstexact
$$
\eqalignno{
b &= -{1 \over \eta}(\bar{\lambda}\gamma^{ab}\bar{\lambda})P_{a}(\lambda\gamma_{b}d) + {1\over 2\eta}(\bar{\lambda}\gamma^{bc}\bar{\lambda})P^{a}(\lambda\gamma_{abc}d) + O(r) & \cr
&= {1\over 2\eta}(\bar{\lambda}\gamma_{bc}\bar{\lambda})(\lambda\gamma^{bc}\gamma^{a}d)P_{a} + O(r) & \bbbrstexact
}
$$
The constraint algebra $\{d_{\alpha}, d_{\beta}\} = -(\gamma^{a})_{\alpha\beta}P_{a}$, then shows our claim.

\medskip 
Finally, after expanding eqn. \physicalbghostgi, and do some algebraic manipulations, one finds that
\eqnn \bghostexpanded
$$
\eqalignno{
b =& {1\over 2\eta}(\bar{\lambda}\gamma_{bc}\bar{\lambda})(\lambda\gamma^{bc}\gamma^{a}d)P_{a} + {4 \over \eta ^{2}}L^{(1)}_{bc,de}(\lambda\gamma^{bcdfa}\lambda)P_{a}N^{e}{}_{f} + {4 \over 3\eta^2}  L^{(1)}_{bf,cf}(\lambda\gamma^{abcde}\lambda)P_{a}N_{de} & \cr
& + {4\over 3 \eta^2}L^{(1)}_{bc,da}(\lambda\gamma^{bcdef}\lambda)P^{a}N_{ef} + {2\over \eta^2}L^{(1)}_{ab,cd}(\lambda\gamma^{a}d)(\lambda\gamma^{bcd}d) - {16\over \eta^3}L^{(2)}_{ab,cd,ef}(\lambda\gamma^{bcdeg}\lambda)N^{f}{}_{g}(\lambda\gamma^{a}d)& \cr
& - {8\over \eta^{3}}L^{(2)}_{ag,cd,bg}(\lambda\gamma^{bcdef}\lambda)N_{ef}(\lambda\gamma^{a}d) -{4\over 3\eta^{3}}L^{(2)}_{ij,ab,cd}(\lambda\gamma^{aijkl}\lambda)(\lambda\gamma^{bcd}d)N_{kl} & \cr
&- {16\over 3\eta^4}L^{(3)}_{ig,ab,cd,jg}(\lambda\gamma^{bcdef}\lambda)(\lambda\gamma^{aijkl}\lambda)N_{kl}N_{ef} - {32\over 3\eta^{4}}L^{(3)}_{ij,ab,cd,ef}(\lambda\gamma^{aijkl}\lambda)(\lambda\gamma^{bcdeg}\lambda)N_{kl}N^{f}{}_{g} & \cr
& & \bghostexpanded
}
$$
which, by simple inspection, coincides with the original expression displayed in \bghost.

\medskip 
\noindent Next we use the new form for the b-ghost, eqn. \physicalbghost, to calculate different quantities relevant to the computation of scattering amplitudes in pure spinor worldline and field theory.

\newsec Some applications

\seclab \secfive

\noindent 

\subsec The ghost number two vertex operator

\subseclab \secfiveone

\noindent
In \maxnotesworldline, a ghost number two vertex operator was defined by letting a non-Lorentz covariant b-ghost act on the ghost number three vertex operator $\Psi$. The result was remarkably shown to be independent of non-minimal variables up to BRST-exact terms. Here, we define the ghost number two vertex operator following the same prescription of \maxnotesworldline
\eqnn \ghostnumbertwovo
$$
\eqalignno{
U^{(2)} &= \{b, \Psi\} & \ghostnumbertwovo
}
$$
Notice that this computation would be pretty complicated to carry out by using the original or simplified expressions for the b-ghost, eqns. \bghost\ and \simplifiedbghost. However, the use of the physical operators discussed in previous section provides a simple and efficient treatment to the problem. Concretely, eqns. \chataonpsi, \phihataonpsi, \phihatalphaonpsi\ yield
\eqnn \bonpsi
$$
\eqalignno{
U^{(2)} =& {3\over 2}P^{a}C_{a} + {3\over 2}(\lambda\gamma^{a}d)\Phi_{a} - {1\over 2}N^{ab}(\lambda\gamma_{ab}\Phi) + Q\bigg[-{3\over 2}P^{a}\rho_{a} - {3\over 2}(\lambda\gamma^{a}d)s_{a} - {3\over 2}(\lambda\gamma^{a}w)(\lambda\gamma_{a}\kappa)\bigg]  & \cr 
& + {3\over 2}{\bf C}^{a}\partial_{a}\Psi + {3\over 2}{\bf \Phi}^{a}(\lambda\gamma_{a}D)\Psi + {9 \over 2}(\lambda\gamma^{a}C)(\lambda\gamma_{a}{\bf \Phi}) & \bonpsi
}
$$
On the other hand, the use of the 11D supergravity equations of motion allows us to show the following identity
\eqnn \bonpsisecondline
$$
\eqalignno{
{3\over 2}{\bf C}^{a}\partial_{a}\Psi + {3\over 2}{\bf \Phi}^{a}(\lambda\gamma_{a}D)\Psi + {9 \over 2}(\lambda\gamma^{a}C)(\lambda\gamma_{a}{\bf \Phi}) =& 
%-{9\over 2}(\lambda\gamma^{ab}\lambda){\bf C}_{a}\Phi_{b} + {9 \over 2}{\bf C}^{a}QC_{a} - {9\over 2}{\bf \Phi}^{a}Q[(\lambda\gamma_{a}C)] & \cr 
%& - {9 \over 2}{\bf \Phi}^{a}(\lambda\gamma^{ba}\lambda)C_{b} + {9 \over 2}(\lambda\gamma^{a}C)(\lambda\gamma_{a}{\bf \Phi}) & \cr 
%=& -{9\over 2}(\lambda\gamma^{ab}\lambda){\bf C}_{a}\Phi_{b} +Q[-{9\over 2}{\bf C}^{a}C_{a}] + {3\over 2}P^{a}C_{a} & \cr 
%& + {9 \over 2}(\lambda\gamma^{ab}\lambda){\bf \Phi}_{b}C_{a} + Q[-{9 \over 2}{\bf \Phi}^{a}(\lambda\gamma_{a}C)] + {9\over 2}(\lambda\gamma^{a}{\bf \Phi})(\lambda\gamma_{a}C) & \cr
%& - {9\over 2}{\bf \Phi}_{b}(\lambda\gamma^{ab}\lambda)C_{a} + {9 \over 2}(\lambda\gamma^{a}C)(\lambda\gamma_{a}{\bf \Phi}) & \cr 
%=& -{9\over 2}(\lambda\gamma^{ab}\lambda){\bf C}_{a}\Phi_{b} + {3\over 2}P^{a}C_{a} + Q[-{9\over 2}{\bf C}^{a}C_{a} -{9 \over 2}{\bf \Phi}^{a}(\lambda\gamma_{a}C)] & \cr 
{3\over 2}P^{a}C_{a} + {3 \over 2}(\lambda\gamma^{a}d)\Phi_{a} - {3\over 2}(\lambda\gamma^{a}w)(\lambda\gamma_{a}\Phi) & \cr
& + Q\bigg[-{9\over 2}{\bf C}^{a}C_{a} -{9 \over 2}{\bf \Phi}^{a}(\lambda\gamma_{a}C) + {9 \over 2}(\lambda\gamma^{a}{\bf C})\Phi_{a}\bigg] & \cr 
& & \bonpsisecondline
}
$$
In this manner, one learns that
\eqnn \bonpsifinal
$$
\eqalignno{
U^{(2)} =& \,\,3 P^{a}C_{a} + 3 (\lambda\gamma^{a}d)\Phi_{a} - 3(\lambda\gamma^{a}w)(\lambda\gamma_{a}\Phi) + Q\bigg[-{3\over 2}P^{a}\rho_{a} - {3\over 2}(\lambda\gamma^{a}d)s_{a} - {3\over 2}(\lambda\gamma^{a}w)(\lambda\gamma_{a}\kappa) & \cr 
& -{9\over 2}{\bf C}^{a}C_{a} -{9 \over 2}{\bf \Phi}^{a}(\lambda\gamma_{a}C) + {9 \over 2}(\lambda\gamma^{a}{\bf C})\Phi_{a}\bigg] & \bonpsifinal
}
$$
The vertex \bonpsifinal\ is manifestly Lorentz covariant and invariant under the pure spinor constraint, and its non-BRST-exact piece is remarkably independent of non-minimal variables. Such a sector matches the vertex found in \maxnotesworldline\ using the Y-formalism in 11D. 
%or, equivalently
%\eqnn \bonpsifinal
%$$
%\eqalignno{
%U^{(2)} =& 3 P^{a}C_{a} + 3 (\lambda\gamma^{a}d)\Phi_{a} - N^{ab}(\lambda\gamma_{ab}\Phi) + Q\bigg[-{3\over 2}P^{a}\rho_{a} - {3\over 2}(\lambda\gamma^{a}d)s_{a} - {1\over 2}N^{ab}(\lambda\gamma_{ab}\kappa) & \cr 
%& -{9\over 2}{\bf C}^{a}C_{a} -{9 \over 2}{\bf \Phi}^{a}(\lambda\gamma_{a}C) + {9 \over 2}(\lambda\gamma^{a}{\bf C})\Phi_{a}\bigg] & \bonpsifinal
%}
%$$

\subsec Generalized Siegel gauge

\subseclab \secfivetwo

\noindent
The maximally supersymmetric theories admitting pure spinor field theory descriptions exhibit a notable symmetry between fields and antifields in a single pure spinor superfield, and thus cannot be quantized by using conventional gauge-fixing techniques. Indeed, it was suggested in \pssreview\ that, in analogy with string field theory, the Siegel gauge $b \Psi = 0$ may be used as a consistent gauge-fixing condition in pure spinor master actions. A slightly modified version, referred to as the generalized Siegel gauge, $b\Psi = Q\Omega$ for some $\Omega$, was used in \maxmaor\ in the context of 10D super-Yang-Mills to show that the scattering amplitudes obtained from the field theory action, match those obtained from CFT techniques in the open superstring \stieberger. 

\medskip
The new expression for the 11D b-ghost \physicalbghost\ will now be used to show that the ghost number three vertex operator $\Psi$ satisfies the generalized Siegel gauge. This easily follows from our results \chataonpsi, \phihataonpsi, \phihatalphaonpsi:
\eqnn \gsiegelgauge
$$
\eqalignno{
b\Psi =& {3\over 2} \partial^{a}C_{a} + {3\over 2}(\lambda\gamma^{a}D)\Phi_{a} + {3\over 2}(\lambda\gamma^{a}D)(\lambda\gamma_{a}\kappa) - {3\over 2}(\lambda\gamma^{a}\partial_{\lambda})(\lambda\gamma_{a}\Phi) - {3\over 2}(\lambda\gamma^{a}\partial_{\lambda})(\lambda\gamma_{a}Q\kappa) & \cr 
&+ Q\bigg[-{3\over 2}\partial^{a}\rho_{a} - {3\over 2}(\lambda\gamma^{a}D)s_{a}\bigg] & \gsiegelgauge
}
$$
Using the transversality of $C_{a\alpha\beta}$, and the linearized 11D supergravity equations of motion (see Appendix A), one then concludes that
\eqnn \gsiegelgaugetwo
$$
\eqalignno{
b\Psi =& 
%{3\over 2} \partial^{a}C_{a} + {3\over 2} (\lambda\gamma^{a}\gamma_{a}\Phi) + {3\over 2}(\lambda\gamma^{a})^{\delta}(\lambda\gamma_{a})_{\alpha}h_{\delta}{}^{\alpha} - {3\over 2}(\lambda\gamma^{a}\partial_{\lambda})(\lambda\gamma_{a}\Phi) & \cr
% + Q\bigg[-{3\over 2}(\lambda\gamma_{a}h^{a}) - {3\over 2}\partial^{a}\rho_{a} - {3\over 2}(\lambda\gamma^{a}D)s_{a} - {3\over 2}(\lambda\gamma^{a}\partial_{\lambda})(\lambda\gamma_{a}\kappa)\bigg] & \cr 
%=&
Q\bigg[-{3\over 2}(\lambda\gamma_{a}h^{a}) - {3\over 2}\partial^{a}\rho_{a} - {3\over 2}(\lambda\gamma^{a}D)s_{a} - {3\over 2}(\lambda\gamma^{a}\partial_{\lambda})(\lambda\gamma_{a}\kappa)\bigg] & \gsiegelgaugetwo
}
$$
as stated.

\subsec The two-particle superfield

\subseclab \secfivethree

\noindent
The pure spinor description of 11D supergravity was introduced by Cederwall in \pssugra. The action is quartic in the pure spinor superfield $\Psi$, and produces the following equation of motion
\eqnn \elevendeom
$$
\eqalignno{
Q\Psi + {\kappa\over 2}(\lambda\gamma_{ab}\lambda){\bf{\Phi}}^{a}\Psi{\bf{\Phi}}^{b}\Psi + {\kappa\over 2}\Psi\{Q,{\bf T}\}\Psi - \kappa^2(\lambda\gamma_{ab}\lambda){\bf T}\Psi{\bf{\Phi}}^{a}\Psi{\bf{\Phi}}^{b}\Psi &= 0 &\elevendeom
}
$$
where ${\bf {\Phi}}^{a}$ is a physical operator introduced in section 3, and ${\bf T}$ is defined as
\eqnn \tdefinition
$$
\eqalignno{
{\bf T} &={32\over 9\eta^3}(\bar{\lambda}\gamma^{ab}\bar{\lambda})(\bar{\lambda}r)(rr)N_{ab}& \tdefinition
}
$$
The use of the perturbiner method allows one to solve eqn. \elevendeom\ in terms of multiparticle superfields. Concretely, the expansion
\eqnn \perturbinerelevend
$$
\eqalignno{
\Psi &= \sum_{{\cal P}} \Psi_{{\cal P}}e^{ik_{{\cal P}}\cdot X}  & \perturbinerelevend
}
$$
where ${\cal P}$ denotes non-empty words $p_{1}p_{2}\ldots p_{m}$, with $p_{1} < p_{2} < \ldots < p_{m}$, and $k_{{\cal P}} = k_{p_{1}} + k_{p_{2}} + \ldots + k_{p_{m}}$, yields the following set of relations:
\eqnn \oneparticle
\eqnn \twoparticle
\eqnn \threeparticle
$$
\eqalignno{
Q\Psi_{p_1} &= 0 &  \oneparticle\cr 
Q\Psi_{p_{1}p_{2}} &= -\kappa (\lambda\gamma_{ab}\lambda){\bf{\Phi}}^{a}\Psi_{p_{1}}{\bf{\Phi}}^{b}\Psi_{p_{2}} - {\kappa \over 2}\Psi_{p_{1}}\{Q, {\bf T}\}\Psi_{p_{2}} - {\kappa \over 2}\Psi_{p_{2}}\{Q, {\bf T}\}\Psi_{p_{1}} & \twoparticle\cr
Q\Psi_{p_{1}p_{2}p_{3}} &= -\sum_{{\cal P} = {\cal Q} U {\cal R}}\kappa \bigg[(\lambda\gamma_{ab}\lambda){\bf{\Phi}}^{a}\Psi_{{\cal Q}}{\bf{\Phi}}^{b}\Psi_{{\cal R}} + {1 \over 2}\Psi_{{\cal Q}}\{Q, {\bf T}\}\Psi_{\cal {R}}\bigg] & \cr
& + \sum_{{\cal P} = {\cal Q} U {\cal R} U {\cal S}}\kappa^2 (\lambda\gamma_{ab}\lambda){\bf T}\Psi_{{\cal Q}}{\bf \Phi}^{a}\Psi_{{\cal R}}{\bf \Phi}^{b}\Psi_{{\cal S}} & \threeparticle\cr
\vdots }
$$
where ${\cal P} = {\cal Q}_{1} U {\cal Q}_{2} U \ldots U {\cal Q}_{s}$, indicates a distribution of the words ${\cal P}$ into the non-empty ordered words ${\cal Q}_{1}$, ${\cal Q}_{2}$, $\ldots$, ${\cal Q}_{s}$. The first equation is nothing but the linearized equation of motion of 11D supergravity obtained from the 11D pure spinor superparticle cohomology. The other equations define the multiparticle superfields of 11D supergravity after removing all BRST-exact terms, as explained in \cederwallequations. To illustrate this, let us study the two-particle superfield. Eqns. \phihataonpsi, \oneparticle\ imply that
\eqnn \twoparticlesimplified
$$
\eqalignno{
Q\tilde{\Psi}_{p_{1}p_{2}} &= -\kappa(\lambda\gamma_{ab}\lambda)\Phi_{p_{1}}^{a}\Phi_{p_{2}}^{b} & \twoparticlesimplified
}
$$
where 
\eqnn \tildepsionetwo
$$
\eqalignno{
\tilde{\Psi}_{p_{1}p_{2}} &= \Psi_{p_{1}p_{2}} +\kappa(\lambda\gamma_{ab}\lambda)s_{p_{1}}^{a}\Phi_{p_{2}}^{b} - \kappa(\lambda\gamma_{ab}\lambda)\Phi_{p_{1}}^{a}s_{p_{2}}^{b}  & \cr 
& + \kappa(\lambda\gamma_{ab}\lambda)s_{p_{1}}^{a}Qs_{p_{2}}^{b} - {\kappa \over 2}\Psi_{p_{1}}{\bf T}\Psi_{p_{2}} - {\kappa \over 2}\Psi_{p_{2}}{\bf T}\Psi_{p_{1}} & \tildepsionetwo
}
$$
Using that $\{Q , b\} = {P^{2}\over 2}$, one finds that
\eqnn \twoparticlesimplified
$$
\eqalignno{
\tilde{\Psi}_{p_{1}p_{2}} &= -{2\kappa \over k_{p_1 p_2}^2}b\bigg[(\lambda\gamma_{ab}\lambda)\Phi_{p_{1}}^{a}\Phi_{p_{2}}^{b}\bigg] & \twoparticlesimplified
}
$$
It is not hard to check that the physical operators studied in section \secfourone, shape the solution of \twoparticlesimplified\ as
\eqnn \twoparticlesuperfieldtwo
$$
\eqalignno{
b[(\lambda\gamma_{ab}\lambda)\Phi_{p_1}^{a}\Phi_{p_2}^{b}] &= \tilde{C}_{p_1 p_2} + Q\Lambda_{p_1 p_2} & \twoparticlesuperfieldtwo
}
$$
where $\Lambda_{p_1 p_2} = -{2 \over k^2_{p_1 p_2}}b(\tilde{C}_{p_1 p_2})$, up to BRST-exact terms, and
\eqnn \twoparticlesuperfieldthree
$$
\eqalignno{
\tilde{C}_{p_1 p_2} =& {1\over 2}\bigg[(\lambda\gamma^{bc}\lambda)h_{p_1,ab}k_{p_2}^{a}\Phi_{p_2,c} + \Omega_{p_1,ab}k_{p_2}^{a}C_{p_2}^{b} - (\lambda\gamma_{b})_{\delta}\lambda^{\alpha}T_{p_1, \alpha a}{}^{\delta}C_{p_2}^{ab} & \cr 
& + (\lambda\gamma^{bc}\lambda)h_{p_2,ab}k_{p_1}^{a}\Phi_{p_1,c} + \Omega_{p_2,ab}k_{p_1}^{a}C_{p_1}^{b} - (\lambda\gamma^{b})_{\delta}\lambda^{\alpha}T_{p_2, \alpha a}{}^{\delta}C_{p_1}^{ab}\bigg] &
\twoparticlesuperfieldthree
}
$$
with $T_{\alpha a}{}^{\delta} = {1\over 36}\bigg[(\gamma^{bcd})_{\alpha}{}^{\delta}H_{abcd} + {1\over 8}(\gamma_{a}{}^{bcde})_{\alpha}{}^{\delta}H_{bcde}\bigg]$ (see \refs{\maxmasoncasaliberkovits,\maxthesis} for details). An easy way of checking this is through the use of the equations of motion listed in Appendix A. For instance, eqns. (A.6), (A.9), (A.14) lead to
\eqnn \multitwoone
$$
\eqalignno{
Q\tilde{C}_{p_1 p_2}  =& {1\over 2}\bigg[(\lambda\gamma^{bc}\lambda)(k_{p_1} \cdot k_{p_2}) \Phi_{p_1,b}\Phi_{p_2,c} - (\lambda\gamma_{b})_{\delta}\lambda^{\alpha}T_{p_1,\alpha a}{}^{\delta} k_{p_2}^{[a}C_{p_2}^{b]}& \cr 
& + Q\bigg[-(\lambda\gamma_{b})_{\delta}\lambda^{\alpha}T_{p_1, \alpha a}{}^{\delta}C_{p_2}^{ab}\bigg] + (1  \leftrightarrow 2)\bigg] & \multitwoone
}
$$
Eqn. (A.15) then requires that
\eqnn \multitwotwo
$$
\eqalignno{
Q\tilde{C}_{p_1 p_2}  =& {1\over 2}\bigg[(\lambda\gamma^{bc}\lambda)(k_{p_1} \cdot k_{p_2}) \Phi_{p_1,b}\Phi_{p_2,c} + (\lambda\gamma_{b})_{\delta}\lambda^{\alpha}T_{p_1,\alpha a}{}^{\delta} Q C_{p_2}^{ab} & \cr 
& + (\lambda\gamma_{b})_{\delta}\lambda^{\alpha}T_{p_1,\alpha a}{}^{\delta}[(\lambda\gamma^{[bc}\lambda)h^{a]}{}_{c,p_2} - (\lambda\gamma^{ab})_{\beta}\Phi^{\beta}_{p_2}] & \cr
& + Q\bigg[-(\lambda\gamma_{b})_{\delta}\lambda^{\alpha}T_{p_1, \alpha a}{}^{\delta}C_{p_2}^{ab}\bigg] + (1  \leftrightarrow 2)\bigg] & \cr 
 =&  {1\over 2}\bigg[(\lambda\gamma^{bc}\lambda)(k_{p_1} \cdot k_{p_2}) \Phi_{p_1,b}\Phi_{p_2,c}  - (\lambda\gamma_{b})_{\delta}\lambda^{\alpha}T_{p_1,\alpha a}{}^{\delta} (\lambda\gamma^{ab})_{\beta}\Phi^{\beta}_{p_2} + (1  \leftrightarrow 2)\bigg] & \cr 
 & & \multitwotwo
}
$$
where we used that $(\lambda\gamma_{b})_{\delta}T_{\alpha a}{}^{\delta}\lambda^{\alpha} = {1\over 12}\bigg[ (\lambda\gamma^{de}\lambda)H_{abde} + {1\over 24}(\lambda\gamma_{ab}{}^{cdef}\lambda)H_{cdef}\bigg]$. The Fierz identity $(\lambda\gamma_{ab})_{\alpha}(\lambda\gamma^{abcdef}\lambda) = -24(\lambda\gamma^{[ab})_{\alpha}(\lambda\gamma^{cd]}\lambda)$, then states that
\eqnn \multitwothree
$$
\eqalignno{
Q\tilde{C}_{p_1 p_2} &= (\lambda\gamma^{bc}\lambda)(k_{p_1} \cdot k_{p_2}) \Phi_{p_1,b}\Phi_{p_2,c} & \multitwothree
}
$$

\newsec Discussions

\seclab \secsix

\noindent 
The main result of this paper is the introduction and construction of the 11D physical operators, and the finding of an alternative formula for the 11D b-ghost, which significantly simplifies algebraic computations in pure spinor superspace. As an exemplification of this statement, we were able to show the defining properties: $\{Q , b\} = {P^2\over 2}$, $\{b,b\} = Q\Omega$, in a systematic and quite simple way. Besides, we provided a few useful applications which will be relevant for studying 11D supergravity interactions from the pure spinor perspective. For instance, the two-particle superfield displayed in eqn. \twoparticlesuperfieldthree\ will be substantial for calculating the 4-point amplitude in pure spinor superspace from the perturbiner method applied to the pure spinor 11D supergravity field theory, see eqn. \threeparticle. Higher-order interactions will require a solid understanding of the different properties associated to the physical operators, e.g. (anti)commutation relations, algebraic identities, and so forth. Likewise, this knowledge might potentially be used for studying consistent deformations of 11D supergravity, in analogy with the maximally supersymmetric Born-Infeld action deduced as the only possible deformation of 10D super-Yang-Mills, satisfying the pure spinor master action \psborninfeld. We plan to tackle these open questions in the near future.

\medskip
\noindent It is exciting to see that the simplification of the 10D b-ghost gave rise to the unravelling of a kinematic algebra which automatically realizes the color-kinematics duality when external states are described by Siegel gauge operators. It would be interesting to use the formulae presented in this work, and to investigate which kind of underlying algebraic structure rules the 11D scattering amplitudes when vertex operators satisfy the Siegel gauge condition. Furthermore, the fundamental role of the 10D b-ghost in loop-level superstring scattering amplitudes suggests that multi-loop 11D pure spinor correlators will require the use and efficient manipulation of this operator, task which might effectively be carried out with the ideas developed in this paper.

\medskip
\noindent It is also worthy pointing out that the simplified version of the 10D b-ghost has been found to be related to a twistorial formulation of 10D super-Yang-Mills using pure spinor variables \refs{\tendsupertwistors,\maxdiegoone,\maxdiegotwo}. This framework was showed to be equivalent to the supertwistor description of ambitwistor strings presented in \nmmax. It is tempting to use the formulae introduced in this work for the 11D b-ghost, and propose a new twistor description of 11D supergravity using pure spinors, with possible stringy realizations. We leave these problems and related issues for future work.

\bigskip \noindent{\bf Acknowledgements:} I am grateful to Martin Cederwall for enlightening discussions on topics related to 11D pure spinors. This work was partially funded by the European Research Council under ERC-STG-804286 UNISCAMP, and by the Knut and Alice Wallenberg Foundation under grant KAW 2018.0162 (Exploring a Web of Gravitational Theories through Gauge-Theory Methods).

%OS is indebted to Song He for insightful discussions and collaboration on related topics. 

%%%%%%%%%%%%%%%%%%%%%%%%%%%%%%%%%%%
%%%%%%%%%%%%%%%%%%%%%%%%%%%%%%%%%%%
%%%%%%%%%%%%%%%%%%%%%%%%%%%%%%%%%%%
\appendix{A}{Linearized 11D supergravity}

\seclab\appendixa

\noindent This Appendix briefly reviews the geometrical construction in superspace which directly reproduces the 11D supergravity equations of motion at linearized order.

\subsec Equations of motion

\noindent Let us first set some notation. We will use capital letters from the beginning/middle of the Latin alphabet to represent tangent/curved superspace indices, and lowercase letters from the beginning (middle) of the Latin/Greek alphabet to denote tangent (curved) space vector/spinor indices. The 11D supergeometry is then defined by the 1-form superfields $E^{A}$ and $\Omega_{B}{}^{C}$, referred to as the vielbein and spin-connection, respectively, and the super-Bianchi identities
\eqnn \elevendgeometry
$$
\eqalignno{
{\cal D}T^{A} = E^{B}R_{B}{}^{A} \ \ &, \ \ {\cal D}R_{A}{}^{B} = 0  & \elevendgeometry\cr
}
$$
where $T^{A} = {\cal D}E^{A}$ is the super-torsion, $R_{A}{}^{B} = {\cal D}\Omega_{A}{}^{B}$ is the super-curvature, and ${\cal D} = E^{A}\nabla_{A}$ is the super-covariant derivative defined to act on the arbitrary tensor ${\cal F}_{A_{1}\ldots A_{m}}{}^{B_{1}\ldots B_{n}}$ as
\eqnn \covariantderivative
$$
\eqalignno{
{\cal D}{\cal F}_{A_{1}\ldots A_{m}}{}^{B_{1}\ldots B_{n}} &= d{\cal F}_{A_{1}\ldots A_{m}}{}^{B_{1}\ldots B_{n}} - \Omega_{A_{1}}{}^{C}{\cal F}_{C A_{2}\ldots A_{m}}{}^{B_{1}\ldots B_{n}} + \ldots + {\cal F}_{A_{1}\ldots A_{m}}{}^{C \ldots B_{n}}\Omega_{C}{}^{B_{1}} + \ldots & \cr
& & \covariantderivative
}
$$
and $d$ is the ordinary exterior derivative. Eqns. \elevendgeometry\ imply the familiar relations
\eqnn \torsiondef
\eqnn \curvaturedef
$$
\eqalignno{
[\nabla_{A}, \nabla_{B}\} &= - T_{AB}{}^{C}\nabla_{C} - 2\Omega_{[AB\}}{}^{C}\nabla_{C} , & \torsiondef\cr
R_{AB,C}{}^{D} &= 2\nabla_{[A}\Omega_{B\}C}{}^{D} + T_{AB}{}^{F}\Omega_{FC}{}^{D} + \Omega_{[AB\}}{}^{F}\Omega_{FC}{}^{D} & \curvaturedef
}
$$
where $[ \, ,\,\}$ means graded commutator. The spectrum of 11D supergravity contains a 3-form gauge field which can be promoted to the 3-form superfield $F = E^{C}E^{B}E^{A}F_{ABC}$, satisfying the gauge transformation $\delta F = d L$, for any 2-form superfield $L$. Its field strength takes the form $G = dF$, and it satisfies the Bianchi identity $d G = 0$. In order to describe linearized 11D supergravity, one first writes the covariant derivative $\nabla_{A} = E_{A}{}^{M}\partial_{M}$ at linear order as
\eqnn \introducingh
$$
\eqalignno{
\nabla_{A} &= D_{A} - h_{A}{}^{B}D_{B}& \introducingh 
}
$$
where $D_{A} = \hat{E}_{A}{}^{M}\partial_{M}$, $h_{A}{}^{B} = \hat{E}_{A}{}^{M}E_{M}^{(1)B} = -E^{(1)M}_{A}\hat{E}_{M}{}^{B}$, ($\hat{E}_{A}{}^{M}$, $\hat{E}_{M}{}^{B}$) are the background values of the vielbeins, and ($E_{A}^{(1)M}$, $E_{M}^{(1)A}$) are their corresponding first order perturbations. Additionally, one imposes the conventional constraints $T_{\alpha\beta}{}^{\delta} = T_{a\alpha}{}^{c} = T_{ab}{}^{c} = G_{\alpha\beta\delta\epsilon} = G_{a\alpha\beta\delta} = G_{abc\alpha} = 0$, and the dynamical contraints $T_{\alpha\beta}{}^{a} = (\gamma^{a})_{\alpha\beta}$, $G_{\alpha\beta ab} = (\gamma_{ab})_{\alpha\beta}$. After plugging \introducingh\ into eqn. \torsiondef, one obtains the following set of equations of motion \refs{\maxmasoncasaliberkovits,\maxthesis}
\eqnn \eomone
\eqnn \eomtwo
\eqnn \eomthree
\eqnn \eomfour
\eqnn \eomfive
\eqnn \eomsix
$$
\eqalignno{
2D_{(\alpha}h_{\beta)}{}^{a} - 2h_{(\alpha}{}^{\delta}(\gamma^{a})_{\beta)\delta} + h_{b}{}^{a}(\gamma^{b})_{\alpha\beta} &= 0 & \eomone \cr
2D_{(\alpha}h_{\beta)}{}^{\delta} - 2\Omega_{(\alpha\beta)}{}^{\delta} + (\gamma^{a})_{\alpha\beta}h_{a}{}^{\delta} &= 0 & \eomtwo\cr
\partial_{a}h_{\alpha}{}^{\beta} - D_{\alpha}h_{a}{}^{\beta} - T_{a\alpha}{}^{\beta} - \Omega_{a\alpha}{}^{\beta} &= 0 & \eomthree\cr
\partial_{a}h_{\alpha}{}^{b} - D_{\alpha}h_{a}{}^{b} - h_{a}{}^{\beta}(\gamma^{b})_{\beta\alpha} + \Omega_{\alpha a}{}^{b} &= 0 & \eomfour\cr
\partial_{a}h_{b}{}^{\alpha} - \partial_{b}h_{a}{}^{\alpha} - T_{ab}{}^{\alpha} &= 0 & \eomfive\cr
\partial_{a}h_{b}{}^{c} - \partial_{b}h_{a}{}^{c} - 2\Omega_{{ab}}{}^{c} &= 0 & \eomsix 
}
$$
The equations of motion associated to the components of the linearized version of the 3-form superfield $F$, can directly be deduced from a 4-form superfield $H$ defined from the field strength $G$ as
\eqnn \tensorcapitalh
$$
\eqalignno{
H_{ABCD} &= \hat{E}_{[D}{}^{Q}\hat{E}_{C}{}^{P}\hat{E}_{B}{}^{N}\hat{E}_{A\}}{}^{M}G_{MNPQ} & \tensorcapitalh 
}
$$
which can equivalently be written as $H_{ABCD} = 4D_{[A}C_{BCD\}} + 6\hat{T}_{[AB}{}^{E}C_{ECD\}}$, where $C_{ABC} = \hat{E}_{[C}{}^{P}\hat{E}_{B}{}^{N}\hat{E}_{A\}}{}^{M}F_{MNP}$, and $\hat{T}^{A}$ is the flat space-valued torsion. 
The expansion of \tensorcapitalh\ then yields
\eqnn \eomseven
\eqnn \eomeight
\eqnn \eomnine
\eqnn \eomten
$$
\eqalignno{
4D_{(\alpha}C_{\beta\delta\epsilon)} + 6(\gamma^{a})_{(\alpha\beta}C_{a\delta\epsilon)} &= 0 & \eomseven\cr
 \partial_{a}C_{\alpha\beta\delta} - 3D_{(\alpha}C_{a\beta\delta)} + 3(\gamma^{b})_{(\alpha\beta}C_{ba\delta)}  &= 3(\gamma_{ab})_{(\alpha\beta}h_{\delta)}{}^{b} & \eomeight \cr
2\partial_{[a}C_{b]\alpha\beta} + 2D_{(\alpha}C_{\beta) ab} + (\gamma^{c})_{\alpha\beta}C_{cab} &= 2(\gamma_{[b}{}^{c})_{\alpha\beta}h_{a]c} + 2(\gamma_{ab})_{(\alpha\delta}h_{\beta)}{}^{\delta} & \eomnine\cr
3\partial_{[a}C_{bc]\alpha} - D_{\alpha}C_{abc} & =  3(\gamma_{[ab})_{\alpha\beta}h_{c]}{}^{\beta} & \eomten
}
$$
The defining relations for the physical operators studied in section \secfourone\ can then be easily found from these equations. For instance, after multiplying by $\lambda^{\alpha}\lambda^{\beta}\lambda^{\delta}$, eqn. \eomseven\ implies that
\eqnn \towardspor
$$
\eqalignno{
3QC_{\epsilon} + D_{\epsilon}\Psi &= -3(\lambda\gamma^{a})_{\epsilon}C_{a} & \towardspor
}
$$
where $C_{\epsilon} = \lambda^{\alpha}\lambda^{\beta}C_{\alpha\beta\epsilon}$, $C_{a} = \lambda^{\alpha}\lambda^{\beta}C_{a\alpha\beta}$. Assuming that there exist the linear operators ${\bf C}_{\epsilon}$, ${\bf C}_{a}$ such that their action on the ghost number three vertex operator $\Psi$ are described by the relations: ${\bf C}_{\epsilon}\Psi = C_{\epsilon} + \ldots$, ${\bf C}_{a}\Psi = C_{a} + \ldots$, where $\ldots$ denote shift symmetry terms \psborninfeld, then eqn. \towardspor\ can be written in the operator form
\eqnn \towardspor
$$
\eqalignno{
[Q , {\bf C}_{\epsilon}]  &= -{1\over 3}d_{\epsilon} -(\lambda\gamma^{a})_{\epsilon}{\bf C}_{a} & \towardspor
}
$$
which is exactly the relation displayed in \drchatalpha. Similar arguments follow for the other operators.

\appendix{B}{11D Pure spinor projector}

\seclab\appendixb

\noindent The 11D pure spinor projector $M_{\alpha}{}^{\beta}$ was originally introduced in \maximalloopcederwall, and shown to be given by
\eqnn \elevenprojectorcederwall
$$
\eqalignno{
M_{\alpha}{}^{\beta} &= \delta_{\alpha}^{\beta} - {1\over 4\alpha}(\bar{\lambda}\gamma_{c})^{\beta}(\lambda\gamma^{c})_{\alpha} - {1\over 2\eta\alpha}(\bar{\lambda}\gamma_{a})^{\beta}(\lambda\gamma^{ab}\lambda)(\bar{\lambda}\gamma_{cb}\bar{\lambda})(\lambda\gamma^{c})_{\alpha} + {1\over 8\alpha}(\bar{\lambda}\gamma_{cd})^{\beta}(\lambda\gamma^{cd})_{\alpha} &\cr
& + {1\over 8\eta\alpha}(\bar{\lambda}\gamma_{ab})^{\beta}(\bar{\lambda}\gamma_{cd}\bar{\lambda})(\lambda\gamma^{ab}\lambda)(\lambda\gamma^{cd})_{\alpha} -{1\over 2\eta\alpha}(\bar{\lambda}\gamma_{ac})^{\beta}(\bar{\lambda}\gamma_{bd}\bar{\lambda})(\lambda\gamma^{ab}\lambda)(\lambda\gamma^{cd})_{\alpha} & \elevenprojectorcederwall
%K_{\alpha}{}^{\beta} =& \delta_{\alpha}^{\beta} - {1\over 4 \alpha}(\bar{\lambda}\Gamma_{a})^{\beta}(\lambda\Gamma^{a})_{\alpha} - {1\over2\eta \alpha}(\bar{\lambda}\Gamma_{a})^{\beta}(\lambda\Gamma^{ab}\lambda)(\bar{\lambda}\Gamma_{cb}\bar{\lambda})(\lambda\Gamma^{c})_{\alpha}\cr
%& + {3\over 4\eta \alpha}(\bar{\lambda}\Gamma_{[ab})^{\beta}(\bar{\lambda}\Gamma_{cd]}\bar{\lambda})(\lambda\Gamma^{ab}\lambda)(\lambda\Gamma^{cd})_{\alpha} & 
}
$$
%The first two lambdabars multiplying the gamma matrices with free indices have downstairs indices, and the last one has an upstairs index.
where $\alpha = \lambda\bar{\lambda}$. This expression can be rewritten in the more convenient way
\eqnn \rewritingprojector
$$
\eqalignno{
M_{\alpha}{}^{\beta} &= \delta_{\alpha}^{\beta} - {1\over 4\alpha}(\bar{\lambda}\gamma_{c})^{\beta}(\lambda\gamma^{c})_{\alpha} - {1\over 2\eta\alpha}(\bar{\lambda}\gamma_{a})^{\beta}(\lambda\gamma^{ab}\lambda)(\bar{\lambda}\gamma_{cb}\bar{\lambda})(\lambda\gamma^{c})_{\alpha} & \cr
& + {1\over 8\eta\alpha}(\bar{\lambda}\gamma_{ab})^{\beta}(\bar{\lambda}\gamma_{cd}\bar{\lambda})(\lambda\gamma^{abcde}\lambda)(\lambda\gamma_{e})_{\alpha} & \rewritingprojector
}
$$
where we used the Fierz identity $(\gamma^{[ab})_{(\delta\epsilon}(\gamma^{cd]})_{\mu)\alpha} = -{1\over 6}(\gamma_{k})_{(\delta\epsilon}(\gamma^{abcdk})_{\mu)\alpha} - {1\over 6}(\gamma^{abcdk})_{(\delta\epsilon}(\gamma_{k})_{\mu)\alpha}$. The application of the familiar 11D identity $(\gamma^{ab})_{(\alpha\beta}(\gamma_{b})_{\delta\epsilon)} = 0$, and \maxnotesworldline
\eqnn \myidentity
$$
\eqalignno{
(\gamma^{ab})_{\alpha}{}^{\beta}(\gamma_{ab})_{\delta}{}^{\epsilon} = 2(\gamma^{a})_{\alpha}{}^{\beta}(\gamma_{a})_{\delta}{}^{\epsilon} + 4(\gamma^{a})_{\alpha}{}^{\epsilon}(\gamma_{a})_{\delta}{}^{\beta} + 4(\gamma^{a})_{\alpha\delta}(\gamma_{a})^{\epsilon\beta} - 4\delta^{\epsilon}_{\alpha}\delta_{\delta}^{\beta} + 4C_{\alpha\delta}C^{\epsilon\beta} & \myidentity
}
$$
imply the following useful relations
\eqnn \relationsidentities
$$
\eqalignno{
{1\over 8\eta\alpha}(\bar{\lambda}\gamma_{ab}w)(\lambda\gamma^{abcde}\lambda)(\bar{\lambda}\gamma_{cd}\bar{\lambda})(\lambda\gamma_{e})_{\alpha} &= {1\over 8\eta\alpha}(\lambda\gamma_{ab}w)(\bar{\lambda}\gamma^{ab}\gamma^{cde}\lambda)(\bar{\lambda}\gamma_{cd}\bar{\lambda})(\lambda\gamma_{e})_{\alpha} + {1\over 4\alpha}(\bar{\lambda}\gamma_{a}w)(\lambda\gamma^{a})_{\alpha} & \cr 
& - {1\over \eta}(\lambda\gamma_{a}w)(\bar{\lambda}\gamma^{ac}\bar{\lambda})(\lambda\gamma_{c})_{\alpha} - {1\over \eta}(w\gamma^{cde}\lambda)(\bar{\lambda}\gamma_{cd}\bar{\lambda})(\lambda\gamma_{e})_{\alpha} & \cr
-{1\over 2\eta\alpha}(\bar{\lambda}\gamma_{a}w)(\lambda\gamma^{ab}\lambda)(\bar{\lambda}\gamma_{cb}\bar{\lambda})(\lambda\gamma^{c})_{\alpha} &= {1\over \eta\alpha}(\lambda\gamma_{a}\bar{\lambda})(\lambda\gamma^{ab}w)(\bar{\lambda}\gamma_{cb}\bar{\lambda})(\lambda\gamma^{c})_{\alpha} + {1\over \eta}(\lambda\gamma_{a}w)(\bar{\lambda}\gamma^{ac}\bar{\lambda})(\lambda\gamma_{c})_{\alpha} & \cr
& & \relationsidentities
}
$$
Therefore, eqn. \rewritingprojector\ takes the equivalent form
\eqnn \projectorfinal
$$
\eqalignno{
M_{\alpha}{}^{\beta} &= \delta_{\alpha}^{\beta} + {1\over \eta}(\lambda\gamma^{cde})^{\beta}(\bar{\lambda}\gamma_{cd}\bar{\lambda})(\lambda\gamma_{e})_{\alpha} + {1\over \eta\alpha}(\lambda\gamma_{a}\bar{\lambda})(\lambda\gamma^{ab})^{\beta}(\bar{\lambda}\gamma_{cb}\bar{\lambda})(\lambda\gamma^{c})_{\alpha} & \cr
& + {1\over 8\eta\alpha}(\lambda\gamma_{ab})^{\beta}(\bar{\lambda}\gamma^{ab}\gamma^{cde}\lambda)(\bar{\lambda}\gamma_{cd}\bar{\lambda})(\lambda\gamma_{e})_{\alpha} & \projectorfinal
}
$$
This equation differs from the 11D projector used in this paper $K_{\alpha}{}^{\beta}$, eqn. \psprojectormine, in the presence of the last two terms. However, these extra terms trivially satisfy the defining properties of a generic projector, and their traces can readily be shown to vanish, meaning they do not contribute to the dimension of pure spinor space. Indeed, if one defines $M_{1,\alpha}{}^{\beta} = {1\over \eta\alpha}(\lambda\gamma_{a}\bar{\lambda})(\lambda\gamma^{ab})^{\beta}(\bar{\lambda}\gamma_{cb}\bar{\lambda})(\lambda\gamma^{c})_{\alpha}$, $M_{2,\alpha}{}^{\beta} = {1\over 8\eta\alpha}(\lambda\gamma_{ab})^{\beta}(\bar{\lambda}\gamma^{ab}\gamma^{cde}\lambda)(\bar{\lambda}\gamma_{cd}\bar{\lambda})(\lambda\gamma_{e})_{\alpha}$, it is not hard to convince oneself that 
\eqnn \ms
$$
\eqalignno{
(\lambda\gamma^{a})_{\beta}M_{1,\alpha}{}^{\beta} = 0 \ &, \ (\lambda\gamma^{a})_{\beta}M_{2,\alpha}{}^{\beta} = 0  & \cr
M_{1,\alpha}{}^{\alpha} = 0 \ &, \ M_{2,\alpha}{}^{\alpha} = 0 & \ms
}
$$
Thus, the only meaningful information is carried by the first two terms of $M_{\alpha}{}^{\beta}$, namely $K_{\alpha}{}^{\beta}$, which satisfies the properties of an actual projector, as discussed in \propertiesofk.

\subsec Equivalence of eqns. \psprojectormine\ and \psprojectorminecompact

\noindent Now we will show that eqn. \psprojectorminecompact\ is identical to \psprojectormine. Indeed, the use of the Fierz identity $(\gamma_{a})_{(\epsilon\alpha}(\gamma^{abc})_{\delta)\rho} = -(\gamma^{[b})_{(\epsilon\alpha}(\gamma^{c]})_{\delta)\rho} + (\gamma^{[b}{}_{k})_{(\epsilon\alpha}(\gamma^{c]k})_{\delta)\rho} + (\gamma^{bc})_{(\epsilon\alpha}C_{\delta)\rho}$, allows one to state
\eqnn \psidentityone
$$
\eqalignno{
(\lambda\gamma^{a})_{\alpha}(\lambda\gamma^{abc})_{\rho} =& -(\lambda\gamma^{[b})_{\alpha}(\lambda\gamma^{c]})_{\rho} - {1\over 2}(\lambda\gamma^{[b}{}_{k}\lambda)(\gamma^{c]k})_{\alpha\rho} + (\lambda\gamma^{[b}{}_{k})_{\alpha}(\lambda\gamma^{c]k})_{\rho} & \cr 
& - {1\over 2}(\lambda\gamma^{bc}\lambda)C_{\alpha\rho} - (\lambda\gamma^{bc})_{\alpha}\lambda_{\rho} & \psidentityone
}
$$
This can be rewritten in the convenient form
%\eqnn \psidentitytwo
%$$
%\eqalignno{
%(\lambda\gamma^{a})_{\alpha}(\lambda\gamma^{abc})_{\rho} =& -C_{\alpha\rho}(\lambda\gamma^{bc}\lambda) + (\lambda\gamma^{a})_{\rho}(\lambda\gamma^{a}\gamma^{bc})_{\alpha} + (\lambda\gamma^{bk})_{\alpha}(\lambda\gamma^{ck})_{\rho} & \cr & + (\lambda\gamma^{k}\gamma^{c})_{\alpha}(\lambda\gamma^{bk})_{\rho} - (\lambda\gamma^{bc})_{\alpha}\lambda_{\rho} & \psidentitytwo
%}
%$$
%Therefore,
% In this expression, one has on the rhs $\lambda^{\delta}(\gamma^{a})_{\delta\alpha}\lambda_{\kappa}(\gamma^{abc})^{\kappa\beta}, and on the lhs the only lambda with downstairs indices is $\lambda_{\delta}(\gamma^{a})^{\delta\beta}$ in the second term$
\eqnn \psidentitythree
$$
\eqalignno{
(\lambda\gamma_{a})_{\alpha}(\lambda\gamma^{abc})^{\beta} =& -\delta_{\alpha}^{\beta}(\lambda\gamma^{bc}\lambda) + (\lambda\gamma_{a})^{\beta}(\lambda\gamma^{a}\gamma^{bc})_{\alpha} - (\lambda\gamma^{[b}{}_{k})_{\alpha}(\lambda\gamma^{c]k})^{\beta} & \cr
& - (\lambda\gamma_{k}\gamma^{[c})_{\alpha}(\lambda\gamma^{b]k})^{\beta} + (\lambda\gamma^{bc})_{\alpha}\lambda^{\beta} & \psidentitythree
}
$$
Therefore, the projector $K_{\alpha}{}^{\beta}$ in eqn. \psprojectorminecompact, can be cast as
\eqnn \psidentityfour
$$
\eqalignno{
K_{\alpha}{}^{\beta} 
%&= \bigg[(\lambda\gamma^{a})^{\beta}(\lambda\gamma^{a}\gamma^{bc})_{\alpha} - (\lambda\gamma^{[bk})_{\alpha}(\lambda\gamma^{c]k})^{\beta}  - (\lambda\gamma^{k}\gamma^{[c})_{\alpha}(\lambda\gamma^{b]k})^{\beta} + (\lambda\gamma^{bc})_{\alpha}\lambda^{\beta}\bigg](\bar{\lambda}\gamma^{bc}\bar{\lambda}) & \cr 
&= {1\over \eta }\bigg[(\lambda\gamma_{a})^{\beta}(\lambda\gamma^{a}\gamma^{bc})_{\alpha} - \lambda_{\alpha}(\lambda\gamma^{bc})^{\beta} + (\lambda\gamma^{bc})_{\alpha}\lambda^{\beta} - 2(\lambda\gamma^{[b}{}_{k})_{\alpha}(\lambda\gamma^{c]k})^{\beta}\bigg](\bar{\lambda}\gamma_{bc}\bar{\lambda}) & \cr 
& & \psidentityfour
}
$$
Using that $(\lambda\gamma^{k})^{\beta}(\lambda\gamma_{k})_{\epsilon} = -{1 \over 6}(\lambda\gamma^{ab})^{\beta}(\lambda\gamma_{ab})_{\epsilon}  - {2\over 3}\lambda_{\epsilon}\lambda^{\beta}$, one arrives at
% Here on the lhs one has in the first term $\lambda_{deta}$, and in the second term $\lambda^{\delta}$
\eqnn \psidentityfive
$$
\eqalignno{
K_{\alpha}{}^{\beta} =& -{1\over 6\eta}(\lambda\gamma^{ab})^{\beta}(\bar{\lambda}\gamma^{cd}\bar{\lambda})(\lambda\gamma_{abcd})_{\alpha} - {4\over 3\eta}(\lambda\gamma^{ck})_{\alpha}(\lambda\gamma_{k}{}^{d})^{\beta}(\bar{\lambda}\gamma_{cd}\bar{\lambda}) - {2\over 3\eta}(\lambda\gamma^{cd})^{\beta}\lambda_{\alpha}(\bar{\lambda}\gamma_{cd}\bar{\lambda}) & \cr 
&+ {1\over 3\eta}\lambda^{\beta}(\lambda\gamma^{cd})_{\alpha}(\bar{\lambda}\gamma_{cd}\bar{\lambda}) & \psidentityfive
}
$$
which coincides with eqn. \psprojectormine.
\listrefs

\bye